 \documentclass[final,5p,times,twocolumn,authoryear]{elsarticle}

\usepackage{amssymb}
\usepackage{lipsum}
\usepackage[lofdepth,lotdepth]{subfig}
\usepackage{gensymb}
\usepackage{amsmath}
\usepackage{hyperref}
\usepackage{lscape}

\newcommand{\be}{\begin{equation}}
\newcommand{\ee}{\end{equation}}
\newcommand{\bary}{\begin{eqnarray}}
\newcommand{\eary}{\end{eqnarray}}

\journal{High Energy Astrophysics}

\begin{document}

\begin{frontmatter}

\title{The unprecedented flaring activities  around Mrk 421 in 2012 and 2013:\\The test for neutrino and UHECR event connection}

\author[first]{Nissim Fraija}
\author[first]{Edilberto Aguilar-Ruiz}
\author[first]{Antonio Galván}
\author[first]{Jose Antonio de Diego Onsurbe}
\author[second,third,fourth]{Maria G. Dainotti}

\affiliation[first]{organization={Instituto de Astronomía, Universidad Nacional Autónoma de México},
            addressline={Circuito Exterior, C.U.}, 
            city={México City},
            postcode={04510}, 
            state={México},
            country={México}}

\affiliation[second]{organization={National Astronomical Observatory of Japan},
            addressline={2-21-1 Osawa},
             city={Mitaka},
             postcode={181-0015},
             state={Tokyo},
             country={Japan}}

\affiliation[third]{organization={Space Science Institute},
            addressline={4765 Walnut St, Suite B},
             city={Boulder},
             postcode={80301},
             state={Colorado},
             country={U.S.A.}}

\affiliation[fourth]{organization={The Graduate University for Advanced Studies, SOKENDAI},
            addressline={Shonan Village},
             city={Hayama},
             postcode={240-0193},
             state={Kanagawa},
             country={Japan}}

%
\begin{abstract}
Since its mission, Fermi Collaboration reported the highest flux observed during July - September 2012 for the BL Lac Markarian 421 (Mrk 421). The integrated flux was eight times greater than the average flux reported in the second Fermi catalog. During April 2013, Mrk 421 exhibited the highest TeV $\gamma$-ray and optical fluxes recorded. The Telescope Array (TA) collaboration reported the arrival of 72 ultra-high-energy cosmic rays (UHECRs), two in temporal and positional coincidence with the flaring activity observed in 2012 and one with the flaring activity in 2013. The IceCube collaboration has reported around  100 neutrino events in the High-Energy Starting Events (HESE) catalog. Although no neutrino track-like event has been associated with this source, a neutrino shower-like event (IC31) is in temporal and positional coincidence with the flare in 2012.  Describing the broadband spectral energy distribution during the flaring activities exhibited in 2012 and 2013 with one- and two-zone lepto-hadronic scenarios and one-zone SSC model, we study a possible correlation between the neutrino event IC31 and the three UHECRs. We estimate the number of neutrino and UHECR events generated from the proposed models, and show that while the neutrino events are low to associate the event IC31 with Mrk 421, the number of UHECRs are similar to those reported by TA collaboration.
\end{abstract}




\end{frontmatter}

\section{Introduction}
At a distance of 134.1 Mpc, the BL Lac object  Markarian 421 \citep[Mrk 421;][]{1991rc3..book.....D, 2005ApJ...635..173S} is one of the closest known and brightest sources in the extragalactic TeV gamma-ray and X-ray sky.  In TeV energies, several campaigns have been carried out  by ground-based Imaging Atmospheric Cherenkov observatories  \citep{1992Natur.358..477P,1996Natur.383..319G,2011ApJ...738...25A, 2002ApJ...575L...9K, 2007ApJ...663..125A, 2005A&A...437...95A, 2002A&A...393...89A, 2003A&A...410..813A,2007ApJ...662..199C,2014APh....54....1A} and air shower arrays (ASAs) \citep{2011ApJ...734..110B,2014ApJ...782..110A,2003ApJ...598..242A}.  In the GeV $\gamma$-ray band, the Large Area Telescope (LAT) onboard the Fermi satellite has observed Mrk 421 since 2009 \citep{2010ApJ...719.1433A}. In X-rays, this source has been observed for more than 14 years by several orbiting satellites  \citep{2010PASJ...62L..55I,2009ApJ...691L..13D, 2012ApJ...759...84N, 2009ATel.2292....1K, 2008ATel.1574....1C,2006ATel..848....1L,2013ATel.4974....1B}.

Different theoretical models have been developed to explain this source's spectral energy distribution (SED) in the quiescent and flaring states. Although theoretical approaches invoking leptonic models to interpret the SED of Mrk 421 have been successful, the presence of protons in the jet has been required to interpret the multi-wavelength observations during the extended quiescent state between 2008 August 5 and 2010 March 12 \citep{2010ApJ...719.1433A} and to describe two atypical TeV flares  without  X-ray counterparts \citep{2005ApJ...630..130B,2011ApJ...738...25A}.  During the extended quiescent state, the GeV $\gamma$-ray flux with a power-law (PL) index of $1.78\pm 0.02$ and average flux of $(7.23\pm 0.16)\times 10^{-8}\, {\rm ph\, cm^{-2}\, s^{-1}}$  was modeled by evoking the synchrotron-proton blazar (SPB) model \citep{2001APh....15..121M,2003APh....18..593M, 2010ApJ...719.1433A}.  The whole SED with two broadband features, one at $\sim$ 1 keV and the other one at hundreds of GeV, was described within the hadronic scenario; the broadband feature at low energies was explained by electron synchrotron radiation, whereas the pion decay products and proton synchrotron explained the high-energy broadband feature \citep{2010ApJ...719.1433A}.  On the other hand, the unusual TeV flares were detected without an increase in the X-ray flux. Firstly, a TeV flare was observed when the X-ray flux exhibited low activity on 2004 January 29 (MJD 53033).   This X-ray flux peaked 1.5 days before the TeV $\gamma$-ray flux.  Secondly,  the higher TeV flux during the night on 2008 May 03 (MJD 54589) was not accompanied by simultaneous X-ray activity.  The last atypical event has been interpreted in the framework of the hadronic scenario \citep{2015APh....71....1F}.

The Telescope Array  (TA) collaboration detected a cluster of 72 ultra-high-energy cosmic rays (UHECRs) \cite{2014ApJ...790L..21A}.  The cluster of events with energies above 57 EeV had a Li-Ma statistical significance of 5.1$\sigma$ during five years of operation. Three UHECRs with energies 57.8, 57.4, and 62.5 EeV were recorded in temporal and positional coincidence with the 2012 and 2013 flaring activities exhibited by Mrk 421.   The field of view of this experiment covers the sky region above -10$^\circ$ of declination, having a good sensitivity in the northern hemisphere \citep{2041-8205-790-2-L21}.   Recently, Kim et al. \cite{Kim:2021Aj} presented the latest results of the hotspot using the most recent data of 11 years, from May of 2008 to May of 2019.   The significance of this hot spot did not increase; rather, it decreased to 3.2$\sigma$. A possible correlation between TA events and the BL Lac  Mrk 421 has been claimed \citep{2014APh....54...61D, 2014ApJ...794..126F, 2015APh....70...54F}.

The IceCube collaboration reported evidence of TeV-PeV extragalactic neutrinos  \cite{2013PhRvL.111b1103A, 2013Sci...342E...1I, 2017arXiv171001191I}. The collaboration reported a total of 102 neutrino events in the High-Energy Starting Events (HESE) catalog \citep{IceCube2020arXiv200109520I, IceCube_2021PhRvD.104b2002A}, and 36 occurrences in the Extremely High Energy (EHE) catalog \citep{IceCube_2016ApJ...833....3A,Icecube_2017ICRC...35.1005H,IceCube_2019ICRC...36.1017S}.  Both catalogs' observed neutrino fluxes yield consistent results with a single PL function with spectral indices $\gamma_{\rm \nu, HESE} = -2.87^{+0.20}_{-0.19}$ and  $\gamma_{\rm \nu, EHE} = -2.28^{+0.08}_{-0.09}$ for HESE and EHE \citep{IceCube_2021PhRvD.104b2002A, Icecube_2017ICRC...35.1005H}, respectively.   The arrival directions of these astrophysical events are compatible with an isotropic distribution.  Although only one neutrino track-like event has been associated with the flaring activity of a known BL Lac object (TXS 0506+056) \citep{IceCube2018Sci...361.1378I, IceCube2018Sci...361..147I}, several astrophysical sources have been proposed as potential candidates \citep[e.g.,][]{2016NatPh..12..807K, 2018A&A...620A.174K, 2019ApJ...880..103G, 2020MNRAS.497..865G, 2021ApJ...912...54R, 2022Sci...378..538I}, and in particular Mrk 421 during the quiescent and flaring states \citep{2015APh....70...54F, 2015IAUS..313..177M, 2016APh....80..115P}. The blazar Mrk 421 exhibited one of the major outbursts \cite{2012ATel.4261....1D} in temporal and positional coincidence with the neutrino event IC31 reported in the HESE catalog \cite{2014arXiv1405.5303A}. This event was detected by IceCube in the third year of observations on 2012 September 06.  The event IC31 had an energy of $42.5^{+5.4}_{-5.7}\,{\rm TeV}$ and a median positional uncertainty of $R_{50}=26^\circ$ centered around the coordinates  ${\rm RA= 146.1^\circ}$ and ${\rm Dec= 78.3}$ (J2000).

On July 16, 2012, ARGO-YBJ recorded a departure of more than four standard deviations from Mrk 421's direction. The most significant flux seen for this source since the start of the Fermi mission was reported by the Fermi team, and it had an integrated flux that was eight times bigger than the average flux published in the second Fermi catalog \citep{2012ApJS..199...31N}.  On April 10, 2013, Mrk 421 exhibited an unheard-of amount of flaring activity in optical wavelengths, X-rays and extremely high energies \citep{2013ATel.4976....1C, 2013ATel.4977....1P, 2014A&A...570A..77P, 2013ATel.4974....1B,2013ATel.4978....1N, 2013ATel.4982....1S, 2017ApJS..232....7F}.

In this work,  we study the gamma-ray flaring activities of Mrk 421 between July and September 2012 and April 2013 and the possible correlations with the high-energy (HE) neutrinos and the UHECR events detected around the flares in 2012 and 2013.  The paper is arranged as follows. In Section 2, we show data analysis and the modelling of the SEDs around the flaring activities of Mrk 421. In Sections 3 and 4, we present  data analysis and discussion of the neutrino and UHECR events, respectively, around these flaring activities. In Section 5, we present a summary.  We use k=$\hbar$=c=1 in natural units and use ``unprimed" and ``prime" to refer quantities in the observer and in the comoving frame, respectively.


\section{Data analysis and Modelling of Spectral Energy Distribution around Flaring activities of Mrk 421}\label{sec:2}

\subsection{Multiwavelength data}

Multiwavelength observations in TeV -  GeV $\gamma$-ray,  X-ray, optical, and radio wavelength bands are plotted in Figure \ref{fig1}.  As follows, we will give a brief introduction about the instruments, energy range, and data used in this work.  

\paragraph{ARGO-YBJ  data.} The Astrophysical Radiation with Ground-based Observatory at YangBaJing (ARGO-YBJ) experiment is an extensive air shower (EAS) array located at an altitude of 4300 a.s.l. at Yang BaJing Cosmic Ray  Laboratory in Tibet, P. R. China. It is sensitive with an energy threshold for primary $\gamma$-rays of $\sim$ 300 GeV.  The detector consists of a carpet ($\sim$ 74$\times$ 78 m$^2$) of resistive plate chambers with $\sim$ 93\% of area effective, surrounded by a partially instrumented area ($\sim$ 20\%) up to $\sim$ 100 $\times$ 110 m$^2$.  The ARGO-YBJ experiment, with a $\sim$ 2  sr FOV, can monitor the sources in the sky with a zenith angle less than 50$^\circ$.  Details on the reduction procedure applied to this data set can be found in \cite{2016ApJS..222....6B}.
\paragraph{VERITAS  data.} The Very Energetic Radiation Imaging Telescope Array System (VERITAS) is a four-12 m diameter IACT located in south Arizona. VERITAS monitored Mrk\,421 between 2007 and 2008,  collecting data for almost 48 hours. Details on data reduction are published in \cite{2011ApJ...738...25A}.
\paragraph{Whipple data.} The Whipple 10m telescope observed Mrk 421 during 2010, accounting for 36 hours. Details about data collected this year are reported in \cite{2015A&A...578A..22A}. 
\paragraph{MAGIC data.} With total exposure of almost 50 hours, MAGIC monitored this BL Lac during the flaring activity of 2013.  Data reduction and analysis above 300 GeV are available \footnote{http://tdx.cat/handle/10803/290265}.   
\paragraph{Fermi-LAT data.} The large area telescope (LAT) onboard the Fermi satellite with a large peak effective area (0.8 m$^2$ for 1 GeV), an energy resolution better than 10\%, and a field of view of about 2.4 sr scans the sky from 20 MeV to more than 300 GeV. The GeV $\gamma$-ray fluxes were obtained between the energy range 0.1-500 GeV using the public database of Fermi-LAT \footnote{http://fermi.gsfc.nasa.gov/ssc/data}.  
\paragraph{Swift-BAT data.} The Burst Alert Telescope (BAT) is a large field of view (1.4 sr) with imaging capabilities in the energy range from 15 to 150 keV \citep{2004ApJ...611.1005G}. The BAT instrument usually observes the sky from 50\% to 80\% of the sky daily.   Swift-BAT provides the daily flux from Mrk 421 at energy 15-50 keV\footnote{Transient monitor results provided by the Swift-BAT team: http://heasarc.gsfc.nasa.gov/docs/swift/results/transients/weak/.}.  
\paragraph{MAXI-GSC data.} The Monitor of the All-sky X-ray (MAXI) mission  Gas Slit Camera \citep[GSC;][]{2009PASJ...61..999M} detector, operating in the 2 - 20 keV range, comprises of 12 one-dimensional position-sensitive proportional counters. It began taking data in August 2009. The light curves for specific sources are in three energy bands: 2-4, 4-10, and 10-20 keV, which were introduced to complete the X-ray spectrum. These data are publicly available \footnote{http://maxi.riken.jp/top}. 
\paragraph{RXT-ASM data.}  The All-Sky Monitor \citep[ASM;][]{1996ApJ...469L..33L} on board the XRTE satellite, which consisted of three proportional counters, was sensitive to an energy range of 2 - 12 keV covering the sky every 1.5 hrs. This data set was obtained from the public MIT archive \footnote{http://heasarc.gsfc.nasa.gov/docs/xte/asm\_products.html}.\\
\paragraph{OAN-SPM data.}  The optical R-band data were performed at the  Observatorio Astron\'omico Nacional of San Pedro M\'artir (OAN-SPM) in Baja California (Mexico) 0.84  m  f/15  Ritchey-Chr\'etien  telescope.   The exposure time was 60 sec per frame of Mrk 421\footnote{Additional technical information can be found in  http://haro.astrossp.unam.mx/blazars/instrument/instrument.html}.\\
\paragraph{OVRO data.}  The Owens Valley Radio Observatory (OVRO; \cite{2011ApJS..194...29R}) is a 40 m radio telescope working at 15 GHz with 3 GHz bandwidth.  As part of the blazar monitoring program, this single-dish observed a sample of over 1800 AGNs twice weekly.  The data are publicly available\footnote{http://www.astro.caltech.edu/ovroblazars/}  and are directly used in this paper.\\


\subsection{Light curve and Flaring activities}
\begin{figure*}
\vspace{0.4cm}
{\centering
\resizebox*{1.1\textwidth}{0.9\textheight}
{\includegraphics{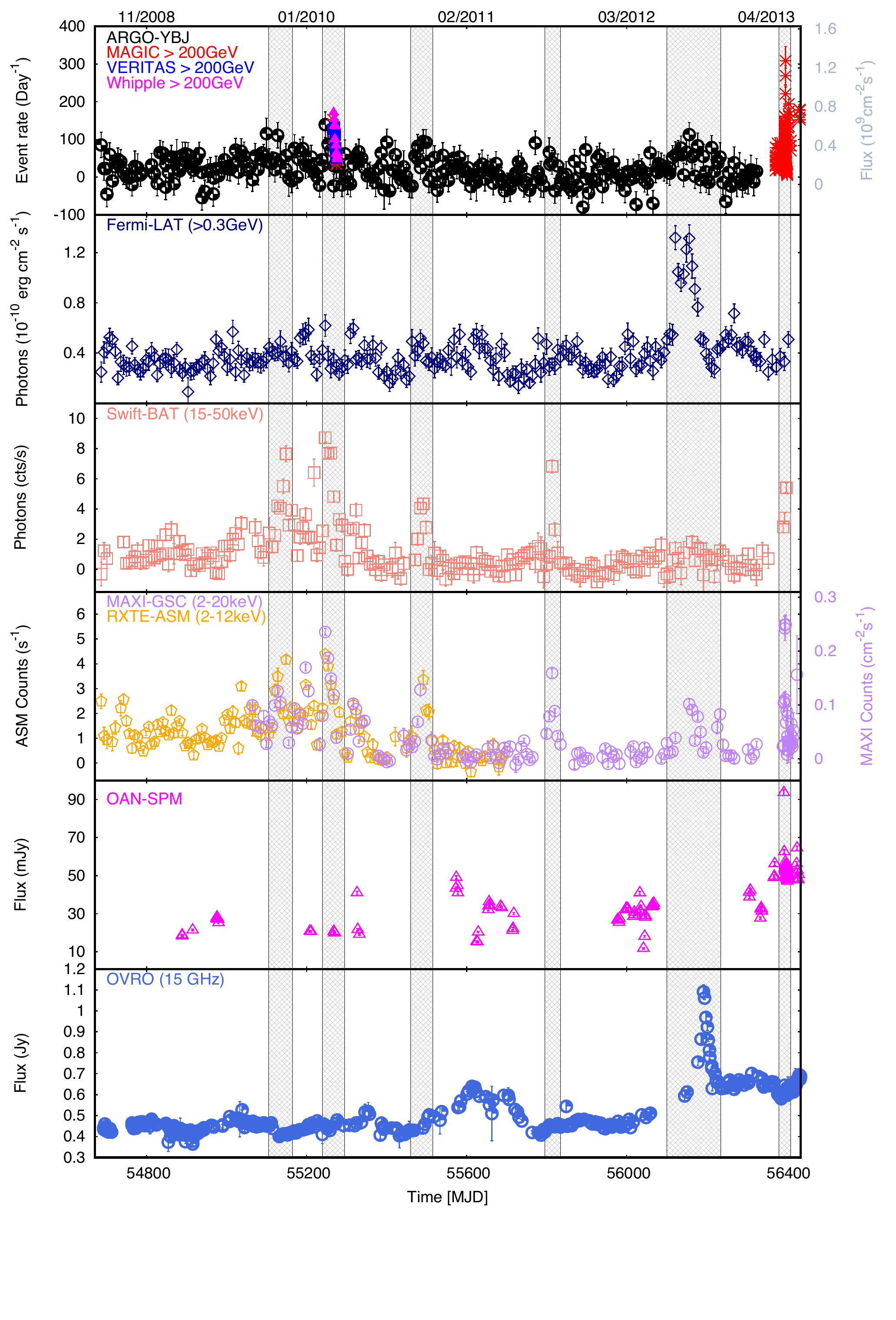}}
}
\caption{Multiwavelength observations of Mrk 421 obtained from ARGO-YBJ, Veritas, Magic, Whipple, Fermi-LAT,  Swift-BAT,  Maxi-GSC, XRT-ASM, OAN-SPM  and  OVRO.  From top to bottom: TeV $\gamma$-ray, GeV $\gamma$-ray,  hard-X ray, soft-X ray, optical and radio bands.}
\label{fig1}
\end{figure*}

Mrk 421 was observed from August 2008 to April 2013 by several instruments in different energy ranges. Bartoli et al. (2016) defined different states of activity for Mrk 421 based on extensive X-ray and GeV $\gamma$-ray flares from August 2008 to February 2013 \cite{2016ApJS..222....6B}. In this work, we add the flaring activity exhibited in April 2013 and the TeV $\gamma$-ray data collected by MAGIC, VERITAS, and Whipple observatories. During the entire observation period (from August 2008 to April 2013), the Mrk 421 light curve exhibited six periods of high activity labeled with F1, F2, F3, F4, F5, and F6, as shown in Figure \ref{fig1}.\\

During the flare F1, the flux began to increase on 2009 November 9, reaching the maximum value on November 12; it decreased to a quasi-steady state on November 14.   The flare F2 started on 2010 February 15 and finished on March 16. The flare F3 lasted around one month from 2010, October 06 to  November 03. The flare F4 occurred in September 2011 and lasted $\sim 7$ days. As we are interested in describing Flares F5 and F6, a more detailed description will be given about these flares.

\paragraph{Flaring activity 2012 (F5).} The longest flare F5 took place on 2012 July 09 and finished on September 16.    On July 16,  ARGO-YBJ reported an excess of 4 standard deviations from a direction consistent with Mrk 421 \citep{2012ATel.4272....1B}. The  Fermi collaboration reported an integrated flux which was eight times greater \cite{2012ATel.4261....1D} than the average flux published in the second Fermi catalog \citep{2012ApJS..199...31N}, making it the highest flux observed for this source since the beginning of the Fermi mission.  This flare was the first long-term GeV - TeV $\gamma$-ray flare from Mrk 421 ever detected by both Fermi-LAT  \citep{2012ATel.4261....1D} and ARGO-YBJ \citep{2012ATel.4272....1B}.   However, neither  MAXI-GSC nor Swift-BAT experiments reported high activities in X-ray bands during this period.\\ 
Using the method of Chi-square $ \chi^2$ minimization as implemented in the ROOT software package \citep{1997NIMPA.389...81B}, we get the integrated flux during 20, 60, and 100 days around the maximum flux for  $\alpha_{\gamma, LAT}=1.75$ \citep{2016ApJS..222....6B}, as reported in Table \ref{table:gammaflux}.

\paragraph{Flaring activity 2013 (F6)}

During 2013 April 10 to 19,  Mrk 421 displayed an unprecedented flaring activity in very-high energies  \citep{2013ATel.4976....1C, 2013ATel.4977....1P}, X-rays \citep{2014A&A...570A..77P, 2013ATel.4974....1B,2013ATel.4978....1N} and optical wavelengths \citep{2013ATel.4982....1S, 2017ApJS..232....7F}.   This blazar displayed the highest optical R-band flux detected, 93.60$\pm$1.53 mJy \citep{2017ApJS..232....7F}.  The multiwavelength light curves  from 2013 April 09 (MJD 56391) to 19 are plotted in Figure \ref{fig1}.\\
The strongest flare was observed from April 10 to 13.  The TeV $\gamma$-ray and optical fluxes were the highest ever recorded for this object \citep{2013ATel.4976....1C}.  In the optical R-band, the flux reached the unprecedented optical luminosity of $L_{\rm R}\simeq9.3 \times 10^{44}$ erg/s.\\

Comparing the XRTE/MAXI and the Swift-BAT observations in Figure \ref{fig1}, we can observe that whereas flares F1, F2, F3, F4, and F6 exhibited a strong correlation in X-ray bands,  in the flare F5, this correlation was absent. For instance, whereas Swift-BAT data did not show activity for the flare F5, MAXI data exhibited low activity for the same period.   


In the framework of the SSC model, photons are radiated via synchrotron emission and up-scatter to higher energies. The low energy emission, from radio to X-rays, is emitted by synchrotron radiation, and the high-energy emission, from MeV to TeV $\gamma$-rays, is produced by inverse Compton scattering. Therefore,  an increase (decrease) in the X-ray flux would be strongly related to the rise (reduction) of the high-energy emission. In the hadronic model, relativistic protons accelerated in the emitting region interact with the seed photons at the lower energy peak or at a different zone. The resulting high-energy photons from these interactions come mainly from  $\pi^0$  decay products, and low-energy components come from secondary $e^\pm$ synchrotron radiation.   In this picture, the high activity in $\gamma$-ray flux could be detected with the absence or low activity in X-ray flux. Therefore, the high activity observed in GeV-TeV $\gamma$-rays with the lack of increased activity in X-rays is challenging to reconcile with the standard SSC, although it has been successfully used to describe the broadband SED of blazars \citep{1996ApJ...461..657B, 1998ApJ...509..608T, 1998MNRAS.301..451G}. In 2012 Mrk 421 exhibited the most significant flare ever observed at radio frequencies \citep{2012ATel.4451....1H}.  Assuming that the GeV $\gamma$-ray and radio are connected,  Hovatta et al. (2015) attempted to model both emissions using the most straightforward theoretical framework \cite{2015MNRAS.448.3121H}.  They found that a simple SSC model can not account for both emissions,  concluding that if physically connected to the preceding $\gamma$-ray flares, it can be reproduced only for a particular choice of parameters. 
 Therefore, the multiwavelength observations of flares F1, F2, F3, F4, and F6 are more consistent with the SSC model and the observations of flare F5 with a lepto-hadronic scenario.

\subsection{Modelling of Spectral Energy Distributions}

The following briefly describes the phenomenological models and radiative processes used to interpret the broadband SEDs of flaring activities in 2012 and 2013.

\subsubsection{One-zone SSC Model}

The one-zone SSC model is the leptonic scenario used to describe the broadband SED of Mrk 421. According to this approach, the magnetic field encloses the electron population accelerated in the emission zone.   Synchrotron radiation emits photons, which are then up-scattered by the parent electron population through inverse Compton scattering up to higher energies. Three power-law functions given by \citep{2011ApJ...736..131A}

\begin{equation}\label{eq:B_SFR}
n'_e (\gamma_e) = N'_{\rm e_1}
    \begin{cases}
        \gamma_e^{-\alpha_{\rm e,1}}\hspace{2.9cm} \gamma_{\rm e, min}<\gamma_e\leq \gamma_{\rm e, c1}
        \\
        \gamma_e^{-\alpha_{\rm e,2}}   \gamma_{\rm e,c1}^{\alpha_{\rm e,2}-\alpha_{\rm e,1}}\hspace{1.4cm} \gamma_{\rm e, c1}<\gamma_e\leq \gamma_{\rm e, c2}
        \\
        \gamma_e^{-\alpha_{\rm e,3}}   \gamma_{\rm e,c1}^{\alpha_{\rm e,2}-\alpha_{\rm e,1}}\gamma_{\rm e, c2}^{\alpha_{\rm e,3}-\alpha_{\rm e,2}}\hspace{0.02cm} \gamma_{\rm e, c2}<\gamma_e\leq \gamma_{\rm e, max},
    \end{cases}
\end{equation}

are used to describe the electron energy distribution.  The parameter $N'_{\rm e_1}$ is the electron number density,  $\alpha_{\rm e, i}$ are the spectral indices for $i=1$, 2 and 3, and $\gamma_{\rm e, j}$ are the electron Lorentz factors for minimum (${\rm j=min}$), breaks (c1 and c2) and maximum (max). 
The electrons in the dissipation zone ($R'_{\rm b}$) permeated by a comoving magnetic field $B'$ cool down following the cooling synchrotron time scale $t'_c=\frac{6\pi m_e}{\sigma_T\,B'^2}\,\gamma^{-1}_e$, with  $\sigma_T$ the Compton cross section and $m_e$ the electron mass.   The minimum variability timescale is given by $\tau_{\rm \nu, min}=(1+z)R'_{\rm b}/\delta_D$ with $\delta_D$ the Doppler factor of the moving emitting region.  The synchrotron spectral breaks are estimated from the electron Lorentz factors as ${\small \epsilon^{\rm syn}_{\rm \gamma, j}(\gamma_{e,j})=\left(\frac{q_e}{\sqrt{8\pi}\,m_e}\right)\,(1+z)^{-1}\,\delta_D\, B'\, \gamma^2_{e,j}}$ with $q_e$ the elementary charge and $z=0.031$ the redshift \citep{2005ApJ...635..173S}.  The maximum Lorentz factor of the electron distribution estimated by equaling the cooling and acceleration timescales is given by $\gamma_{\rm e, max}=(3q_e/\sigma_T B')^{1/2}$. The synchrotron spectrum is derived using Eq. (\ref{eq:B_SFR}) and emissivity $\epsilon_{\gamma}N_\gamma d\epsilon_{\gamma}=\left(-dE_e/dt\right) N_e(E_e)dE_e$ \cite{1986rpa..book.....R, 1994hea2.book.....L}. Fermi-accelerated electrons in the emitting region can upscatter synchrotron photons up to higher energies as 
\be\label{ic}
\epsilon^{\rm ssc}_{\gamma,j}\simeq  \gamma^2_{j} \epsilon^{\rm syn}_{\gamma,j}\,.
\ee
The model of the extragalactic background light (EBL) absorption introduced in \cite{2017A&A...603A..34F} was considered.   The total jet power given a function of energy densities carried by magnetic field ($U_{\rm B}$), electrons ($U_{\rm e}$) and protons ($U_{\rm p}$) is

\be
L_{\rm jet}= L_{\rm B} + L_{\rm e} + L_{\rm p}\,, 
\ee

where $L_{\rm B} \simeq \pi {R'}_b^2 \Gamma^2 U'_{\rm B}$ with $U'_{\rm B}=\frac{B'^2}{8\pi}$, $L_{\rm e} \simeq \pi {R'}_b^2 \Gamma^2 U'_e$ with $U'_{\rm e}=m_{\rm e}n'_{\rm e}\langle\gamma_e\rangle$ and $L_{\rm p}\simeq \pi {R'}_b^2 \Gamma^2 U'_{\rm p}$ with $U'_p= \int \epsilon'_p   n'_p(\gamma_p) d\epsilon'_p$ with $\Gamma$ the Lorentz factor.  The comoving proton density becomes $N'_p=\int   n'_p(\gamma_p) d\epsilon'_p$, where  $n'_p(\gamma_p)$ is given in Eq. \ref{eq_proton_distribution}. It is worth noting that for a charge neutrality condition $N'_p\simeq N'_e$ \citep{2013ApJ...768...54B, 2009ApJ...704...38S, 2011ApJ...736..131A, 2014A&A...562A..12P,2017APh....89...14F}. The one-zone SSC model and the methodology  showed in \cite{2017ApJS..232....7F} will be used to model the  broadband SED.\\ 

\subsubsection{One-zone lepto-hadronic model}

Relativistic protons accelerated in the jet suggest that photo-hadronic interactions must be considered to explain the GeV-TeV $\gamma$-ray emission \citep{2001APh....15..121M}.  In the lepto-hadronic model,  the relativistic electrons injected in the strongly magnetized blob lose energy predominantly by synchrotron radiation, and the resulting photon field serves as targets for in the photo-hadronic interactions.  The low-energy bump in the SED is described by synchrotron radiation, and the high-energy bump is modeled as inverse Compton scattering of synchrotron photons and $\pi^0$ and  $\pi^\pm$ decay products from hadronic interactions.    To describe the broadband SED,  we use the electron and proton distributions characterized by broken and simple PL functions 

{\small
\begin{equation}\label{eq:B_SFR_2}
n'_e(\gamma_e) = N'_{\rm e_2}\,
    \begin{cases}
        \gamma_e^{-\alpha_{\rm e,1}}\hspace{1.7cm} \gamma_{\rm e, min}<\gamma_e\leq \gamma_{\rm e,c}
        \\
        \gamma_{e,c}   \gamma_{\rm e}^{-\alpha_{\rm e,2}}\hspace{0.8cm} \gamma_{\rm e, c}<\gamma_e\leq \gamma_{\rm e, max}\,,
    \end{cases}
\end{equation}
}

and
\be\label{eq_proton_distribution}
n'_p(\gamma_p)=K'_p \gamma_p^{-\alpha_p}\,,
\ee

from $\gamma_{\rm p, min}$ to $\gamma_{\rm p, max}$, respectively.  The parameter $N'_{\rm e_2}$  is the electron number density,  $K'_p$ is the proportionality constant of the proton distribution, $\alpha_{\rm e,2}=\alpha_{\rm e,1}+1$ and $\alpha_p$ are the spectral indices of the electron and proton population, respectively, and $\gamma_{\rm e, min}$, $\gamma_{\rm e, c}$ and $\gamma_{\rm e, max}$ are the electron Lorentz factors for minimum, break and maximum, respectively.  We consider that the electrons and protons are co-accelerated in the same emitting region ($R'_b$) with $\alpha_{\rm e,1}=\alpha_{\rm p}$. {\bf The maximum energy of protons in the comoving frame is $\varepsilon_{p, \rm max}' \simeq  \eta_{\rm acc} \,q_e\, B'\, R'_b$ where $\eta_{\rm acc}$ is the acceleration efficiency factor}.

We suppose that protons are cooled by p$\gamma$ interactions in the inner jet's emission zone. We do not consider proton synchrotron radiation because the total power radiated is much smaller than that power radiated by electrons. The radiated power is inversely proportional to the fourth power of the rest mass \cite{1986rpa..book.....R}.  Charged ($\pi^+$) and neutral ($\pi^0$) pions are created from p$\gamma$ interactions through the following channels   
 \begin{eqnarray}
p\, \gamma &\longrightarrow&
\left\{
\begin{array}{lll}
p\,\pi^{0}\   &&   \mbox{fraction }2/3, \\
n\,  \pi^{+}      &&   \mbox{fraction }1/3\,.\nonumber
\end{array}\right. \\
\end{eqnarray}
Subsequently,  neutral pion decays into photons, $\pi^0\rightarrow \gamma\gamma$,  carrying $20\% (\xi_{\pi^0}=0.2)$ of the proton's energy $E_p$.   The efficiency of the photo-pion processes is \citep{1968PhRvL..21.1016S, PhysRevLett.78.2292}
{\small
\begin{equation}\label{eficiency}
f_{\pi^0} =\frac{R'_b}{2\gamma^2_p}\int\,d\epsilon\,\sigma_\pi(\epsilon)\,\xi_{\pi^0}\,\epsilon\int dx\, x^{-2}\, \frac{dn_\gamma}{d\epsilon_\gamma} (\epsilon_\gamma=x)\,,
\end{equation}
}
where $dn_\gamma/d\epsilon_\gamma$ is the spectrum of seed photons,  $\sigma_\pi(\epsilon_\gamma)$ is the cross section of pion production and $\gamma_p$ is the proton Lorentz factor. Solving the integrals we obtain 
{\small
\bary
f_{\pi^0} \simeq \frac{\sigma_{\rm p\gamma}\,\Delta\epsilon_{\rm res}\,\xi_{\pi^0}\, L_{\rm \gamma}}{4\pi\,\Gamma^2\,R'_b\,\epsilon_{\rm pk,\gamma}\,\epsilon_{\rm res}}
\begin{cases}
\left(\frac{\epsilon^{\pi^0}_{\gamma}}{\epsilon^{\pi^0}_{\gamma,c}}\right)^{\beta_h-1}       &  \epsilon_{\gamma} < \epsilon^{\pi^0}_{\gamma,c}\\
\left(\frac{\epsilon^{\pi^0}_{\gamma}}{\epsilon^{\pi^0}_{\gamma,c}}\right)^{\beta_l-1}                                                                                                                                                                                                            &   \epsilon^{\pi^0}_{\gamma,c} < \epsilon_{\gamma}\,,\\
\end{cases}
\eary
}
where $\beta_h\sim 2.5$ and $\beta_l\sim 1.5$ are the high-energy and low-energy photon indexes, respectively,  $L_{\rm \gamma}$ is the observed luminosity of the seed photons, $\Delta\epsilon_{\rm res}$=0.2 GeV,  $\epsilon_{\rm res}\simeq$ 0.3 GeV, $\epsilon_{\rm \gamma, pk}$ is the energy around the seed photons and  $\epsilon^{\pi^0}_{\gamma,c}$ is the break photo-pion energy given by \cite{2003ApJ...586...79A, 2007Ap&SS.309..407H, 2008PhR...458..173B}
\be
\epsilon^{\pi^0}_{\gamma,c}\simeq 15.94\,{\rm GeV}\, \Gamma^2\, \left(\frac{\epsilon_{\rm \gamma, pk}}{ {\rm MeV}}\right)^{-1}\,.
\label{pgamma}
\ee
It is important to note that for this process is useful to estimate the optical depth and the comoving density of the target photons, which is given by
\be
\tau_{\gamma}\simeq\frac{L_{\rm \gamma}\,\sigma_T}{ 20 \pi R'_b\,\Gamma^3\epsilon_{\rm \gamma, pk}}\,,
\ee
and
\be
n'_{\gamma}\simeq\frac{L_{\rm \gamma}}{4 \pi {R'}^2_b\, \Gamma^3 \epsilon_{\rm \gamma, pk}}\,,
\label{den}
\ee
respectively. The photo-pion spectrum can be estimated  from photons released  in the range $\epsilon^{\pi^0}_\gamma$ to $\epsilon^{\pi^0}_\gamma + d\epsilon^{\pi^0}_\gamma$ by protons in the range   $\varepsilon_p$ and $\varepsilon_p + d\varepsilon_p$ and the photo-pion efficiency. Details of the lepto-hadronic model can be found in \citet{2016ApJ...830...81F}.

\subsubsection{Two-zone lepto-hadronic model}

Recently, \cite{AguilarRuiz_2022MNRAS.512.1557A} proposed a lepto-hadronic model with a two-zone dissipation region. In this scenario, an outer blob \footnote{To express the quantities measured in this blob we mark them using two-prime.} produces the most significant flux in X-ray and GeV-gamma-ray bands by emitting electrons under an SSC model, and the inner blob is responsible for the highest gamma-ray emission via the photopion processes. These photopions result from the interaction of accelerated protons inside this blob and seed photons from a pair-plasma.  In both blobs, the electron distribution is described as a broken PL as given by Eq. \ref{eq:B_SFR}, while the proton distribution in the inner blob is an unbroken PL as Eq. \ref{eq_proton_distribution}.  The relative Lorentz factor gives the relative motion between the inner blob and pair-plasma, which yields
\begin{equation}\label{eq_GammaRel}
\Gamma_{\rm rel} = \Gamma_{\rm i} \Gamma_{\rm pl} \left( 1 - \beta_{\rm i} \beta_{\rm pl} \right) \simeq 0.735 \Gamma_{i} \,,
\end{equation}
where $\Gamma_i ( \Gamma_{\rm pl}\approx 1.05 )$ and $\beta_i \approx 1 (\beta_{\rm pl} = 0.3)$ are the Lorentz factor and the velocity of the inner blob (pair-plasma), respectively.

Photons from the pair-plasma with energy $\varepsilon'_{\rm pl} = 2\Gamma_{\rm rel} \, \varepsilon_{\rm pl}$ are blueshifted within the inner blob's scenario. Similarly, the comoving photon number density measured in the inner blob is
\begin{align}\label{eq_pair_plasma_density}
    n_{\rm pl}' 
    & \simeq \frac{L_{\rm keV}  }{4\Omega_{\rm pl} R^{ 2}_{\rm ph} \beta_{\rm pl } m_e c^3 \, \Gamma_{\rm rel} }   \,  
    \\
    & \approx 6.5 \times 10^{12} \, {\rm cm^{-3}}
    \, \Gamma_{\rm rel}^{-1} 
    \left( \frac{L_{\rm keV}}{\rm 3 \times 10^{44} \, erg \, s^{-1}}\right)
    \,, 
\end{align}
where the luminosity of the pair-plasma $L_{\rm keV}\gtrsim 3 \times 10^{-3}L_{\rm Edd}$ corresponds to a fraction of the Eddington Luminosity $L_{\rm Edd}$, $R_{\rm ph}\sim r_g \approx 10^{14} \rm \, cm$ is the photospheric radius and $\Omega_{\rm pl}\sim 0.2 \pi$ is the solid angle covered by the pair-plasma.  Additionally, photopions generate gamma rays with observed energies greater than

\begin{equation}\label{eq_pion_th_gamma}
    \epsilon_\gamma \simeq 0.16 \, {\rm TeV} \, \mathcal{D}_i \, \Gamma_{\rm rel} \, \left( \frac{\varepsilon_{\rm pl}'}{\rm 100 \,keV}\right)^{-1}\,,
\end{equation}

where $\mathcal{D}_i$ is the Doppler factor of the inner blob. 

We can estimate the synchrotron frequency at the flux peak radiated by the primary electrons using the typical values for the inner blob as $\nu_{e, \rm syn, pk} \approx 16 \, {\rm GHz} \, \left( \frac{\Gamma_i}{3} \right) \left( \frac{B_i'}{100 \, \rm G} \right)^{-3} \left( \frac{R_i'}{10^{14} \rm \, cm} \right)^{-2}$, with $B_i' = (R_o^{''}/R_i') \, B_o^{''} \, \sim 100 \rm \, G ,$
where $B_o^{''}(B_i')$ and $R_o^{''}(R_i')$ correspond to the strength of the magnetic field and blob's radius for the outer (inner) blob, respectively. {\bf In this scenario, the maximum energy of protons in the comoving frame is $\varepsilon_{p, \rm max}' \simeq  \eta_{\rm acc} \,q_e\, B'_{i}\, R'_{i}$}.

Due to the magnetic field is stronger in the inner than in the outer blob, electrons in the inner blob are expected to be cooled more quickly by synchrotron radiation. The X-ray flux created by secondary electrons via synchrotron radiation must be a significant aspect of this scenario $L_{\rm syn, p\pi} \approx (1/4) L_{\gamma,p\pi}$, which has an observed energy peak
\begin{equation}
    \epsilon_{\rm syn,p\pi} \sim 40 \, {\rm keV} \,  \Gamma_{\rm rel}^{2} \, \left(\frac{B_i'}{100 \rm G}\right) \, \left(\frac{\Gamma_i}{3}\right)  \, .
\end{equation}

It is worth noting that photons of the pair-plasma suppress the MeV-GeV emission produced in the inner blob, which is due to the optical depth to photon-photon annihilation $\tau_{\gamma\gamma} \sim 133 \, (n_{\rm pl}'/5 \times 10^{12} \, {\rm cm^{-3}}) \, (R_i'/ 2 \times 10^{14} \, {\rm cm})$.  Therefore, secondary electrons produce a signature between $\rm 40 \, keV \, \Gamma_{\rm rel}^2$ and  $\varepsilon_\gamma \sim 1 \, {\rm MeV} \, \Gamma_{\rm rel} \mathcal{D}_i$. However, gamma-rays at TeV energies, resulting {\bf from} neutral pion decay, can escape from inner blob without significant attenuation. 
Details of this phenomenological model, such as the spectral breaks and the spectra of the secondary electrons and pion decay products, can be found in \citet{AguilarRuiz_2022MNRAS.512.1557A, AguilarRuiz_2023JHEAp..38....1A}.


\subsection{Results and Discussion}

\begin{figure*}
\vspace{0.4cm}
{\centering
\resizebox*{1.05\textwidth}{0.3\textheight}
{\includegraphics{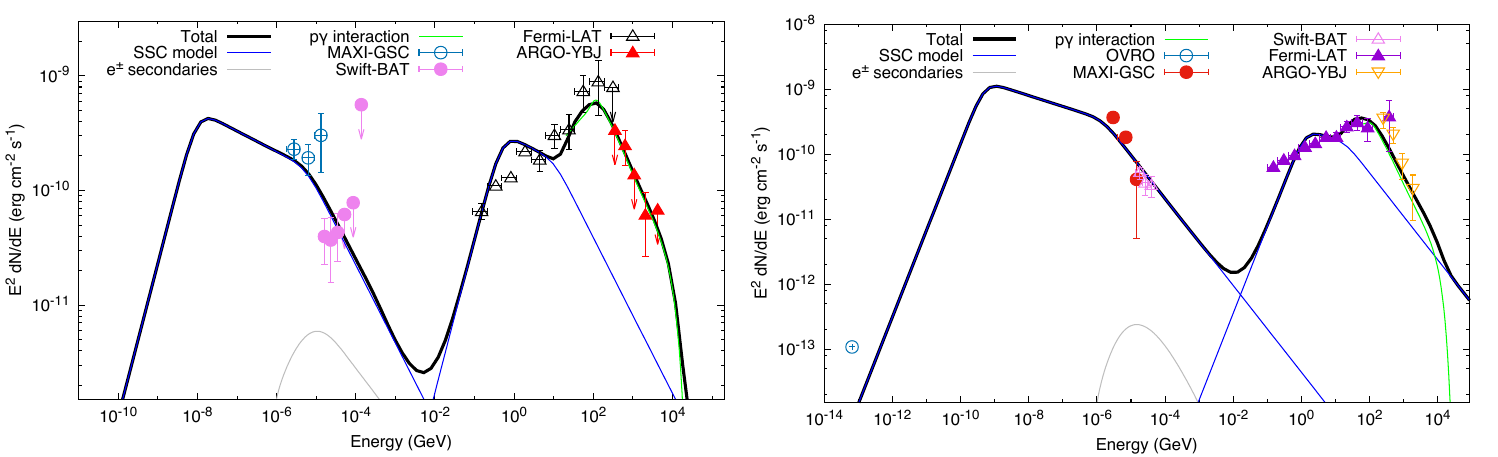}}
}
\caption{The broadband SEDs of Mrk 421 from 2012 July 09 to 21 (left panel) and from 2012 July 22 to September 16 (right panel) with the best-fit model curves for the one-zone lepto-hadronic model. The best-fit parameters are shown in Table \ref{sed_parameters_hadronic_one}.}
\label{SED_flare2012_one}
\end{figure*}
\begin{figure*}
\begin{minipage}[b]{0.5\linewidth}
\centering
\includegraphics[width=\linewidth, height=7cm]{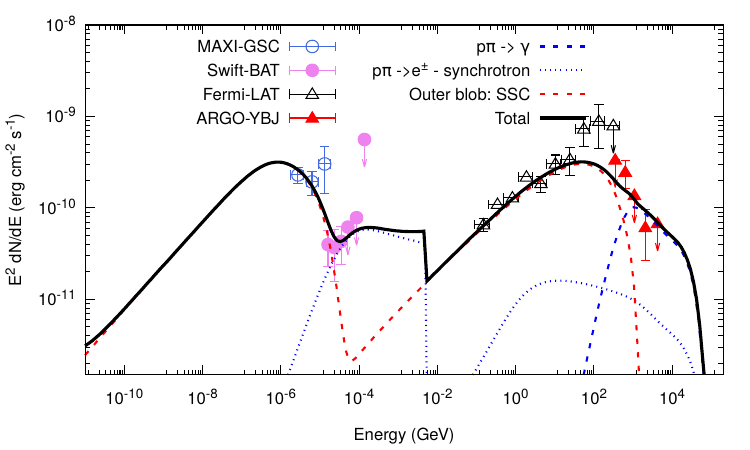}
\end{minipage}\hfill 
\begin{minipage}[b]{0.50\linewidth}
\centering
\includegraphics[width=\linewidth, height=7cm]{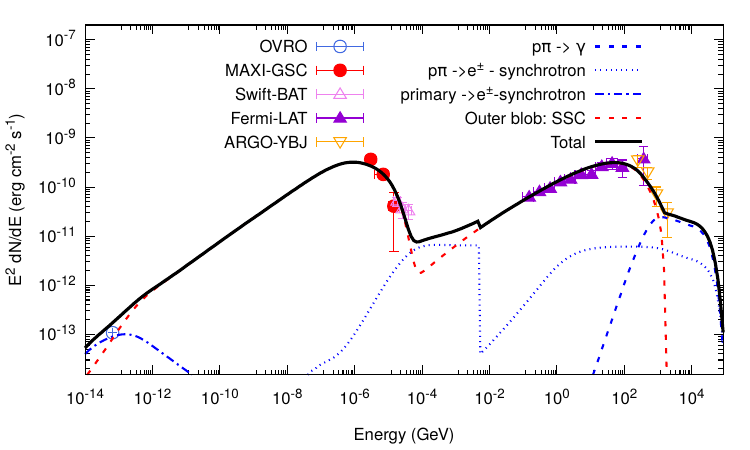} 
\end{minipage}
\caption{The same as Figure \ref{SED_flare2012_one}, but with the best-fit curve for two-zone lepto-hadronic model emission. The best-fit parameters are shown in Table \ref{tab_parameters_twozone_hadronic}.} \label{fig_SED_flare_2012_two}
\end{figure*}
\begin{figure*}
\begin{minipage}[b]{0.5\linewidth}
\centering
\includegraphics[width=\linewidth, height=7cm]{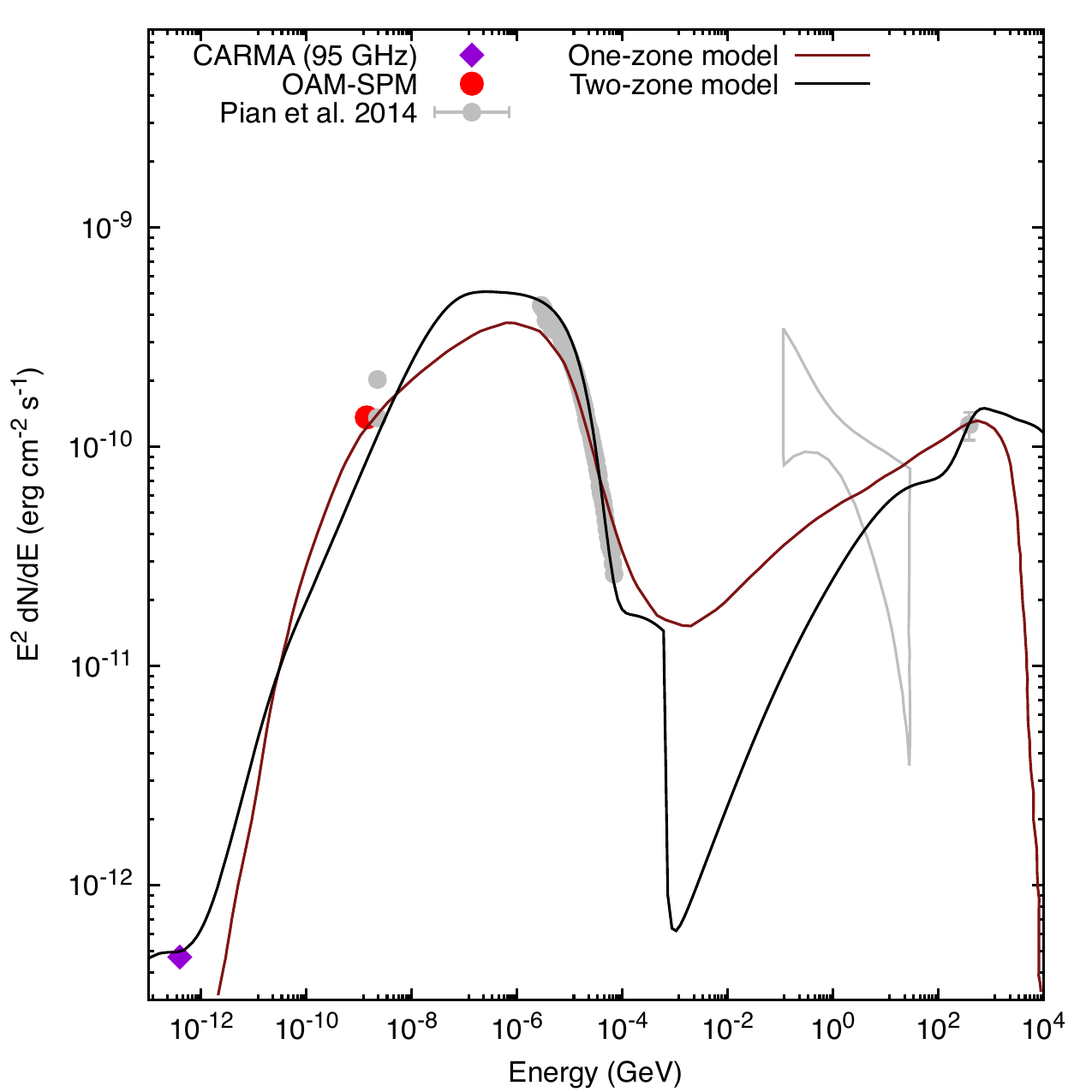}
\end{minipage}\hfill 
\begin{minipage}[b]{0.50\linewidth}
\centering
\includegraphics[width=\linewidth, height=7cm]{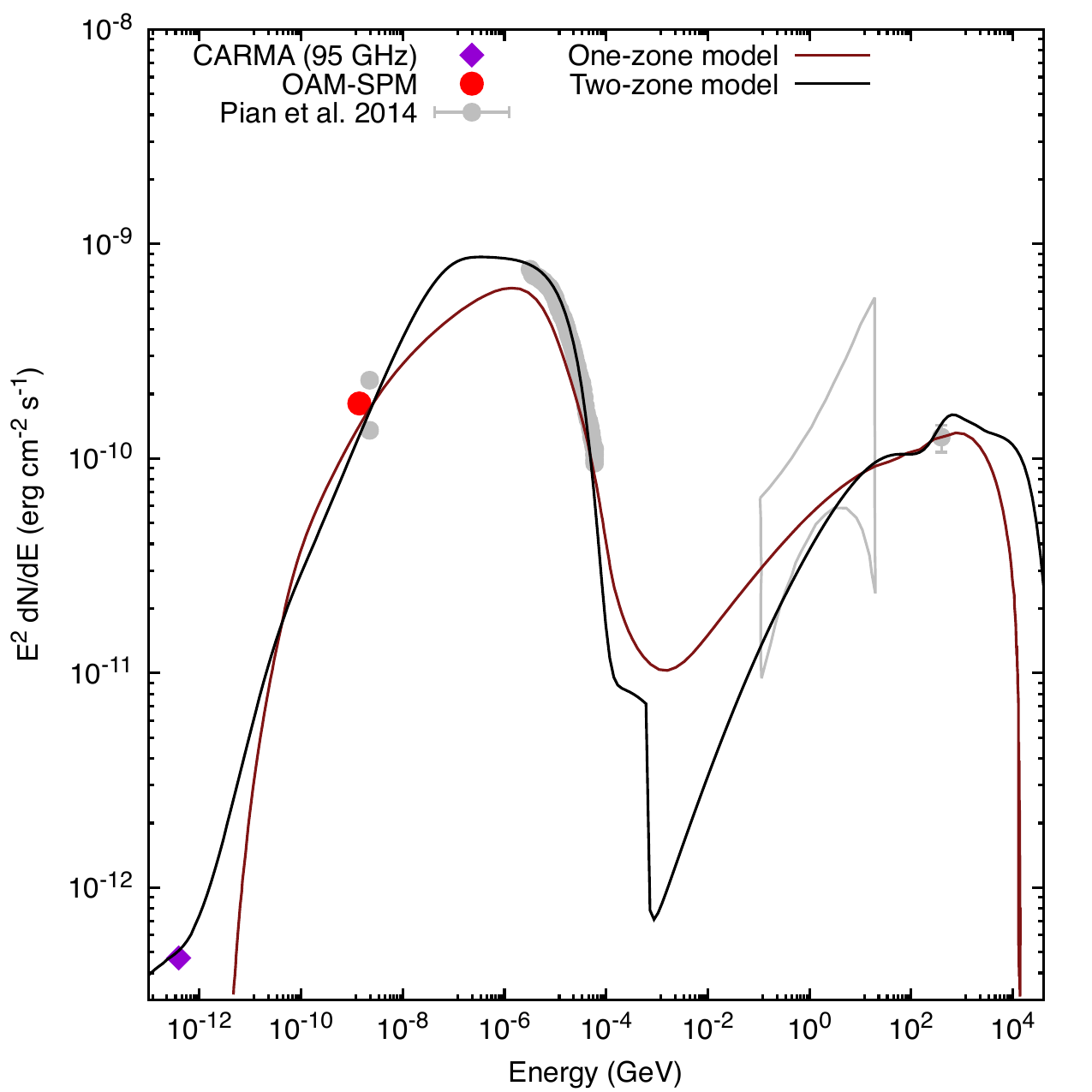} 
\end{minipage}
\caption{\bf The broadband SEDs of Mrk 421 during 2013 April 16 (left panel) and 2013 April 17 (right panel) with the best-fit model curves for the one- and two-zone lepto-hadronic model.  The best-fit parameters are shown in Tables \ref{sed_parameters_hadronic_one} and \ref{tab_parameters_twozone_hadronic} for the one- and two-zone lepto-hadronic model, respectively.} \label{fig_SED_flare_2013_two}
\end{figure*}
\begin{figure}
\vspace{0.4cm}
{\centering
\resizebox*{1.\textwidth}{0.3\textheight}
{\includegraphics{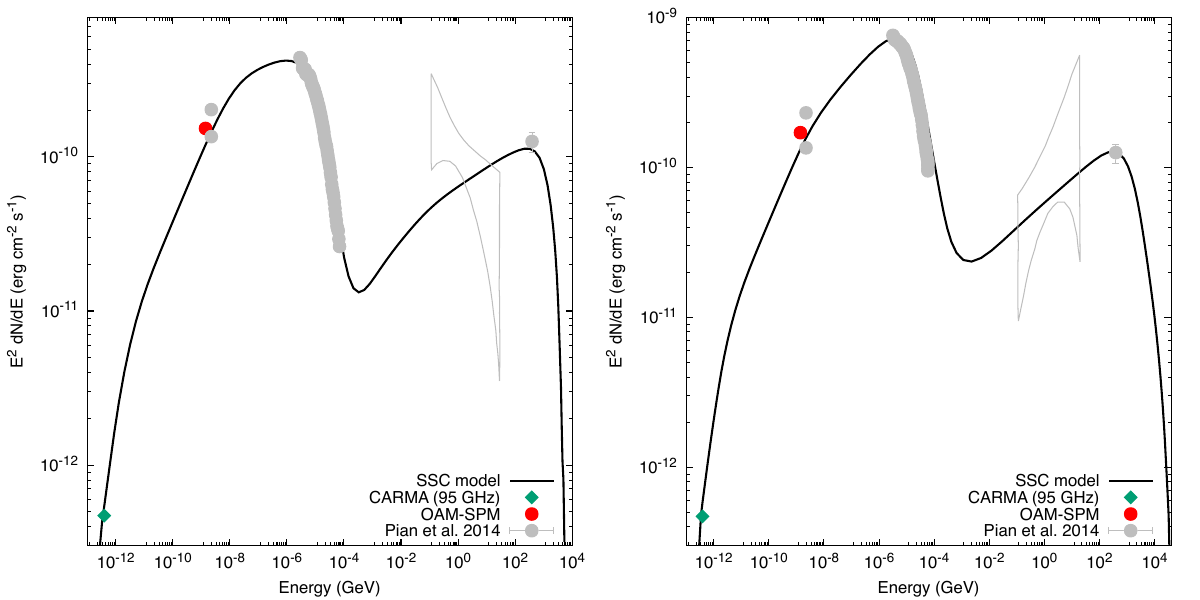}}
}
\caption{The one-zone SSC model was used to describe the SEDs of Mrk\,421 during 2013 April 16 ({\bf left} panel) and 2013 April 17 ({\bf right} panel).  The best-fit parameters are shown in Table \ref{sed_parameters_leptonic}.}
\label{SED_flare2013}
\end{figure}

Figures \ref{SED_flare2012_one} and \ref{fig_SED_flare_2012_two} show the fit of the SEDs of Mrk\,421 from 2012 July 09 to 21 (left-hand panel) and  from 2012 July 22 to September 16 (right-hand panel) using the one- and the two-zone lepto-hadronic model, respectively.   Figure \ref{fig_SED_flare_2013_two} shows the fit of the SEDs of Mrk\,421 on 2013 April 16 (left-hand panel) and 2013 April 17 (right-hand panel) using the one- and the two-zone lepto-hadronic model. The respective best-fit parameters using the one- and two-zone lepto-hadronic models are listed in Tables \ref{sed_parameters_hadronic_one} and \ref{tab_parameters_twozone_hadronic}, respectively.    Figure \ref{SED_flare2013} is similar to Figure \ref{fig_SED_flare_2013_two}, but the one-zone SSC scenario is required.   The best-fit values using the one-zone SSC model are reported in Table \ref{sed_parameters_leptonic}. All Tables show the set of the parameter values obtained and derived.   The complete set of parameters obtained by modelling the SEDs of Mrk 421 with the one-zone lepto-hadronic model lies in the ranges of $14 - 17$, $6.1 - 7.3\times 10^{16}\,{\rm cm}$ and $8-11\,{\rm G}$ for the Doppler factor,  the emitting radius and the magnetic field, respectively, which agrees with values reported in \cite{2011ApJ...736..131A}.


Given the values listed in Tables \ref{sed_parameters_hadronic_one} and \ref{tab_parameters_twozone_hadronic},  we could estimate 
the photon-photon annihilation which could attenuate TeV gamma-ray emission and then creates cascades that show up in X-rays. Similarly, we could estimate Bethe-Heitler pair production. {\bf In this case,  the optical depth of photon-photon annihilation (Bethe-Heitler) for 2012 flare is 
$\tau_{\gamma\gamma}\sim 0.06 $ ($\sim 10^{-2}$)
for one- and 
$\tau_{\gamma\gamma}\sim 80$ ($\sim 13$) 
two-zone lepto-hadronic models}. These estimates show that the effect is subdominant in the effect of Bethe-Heitler is considered, but in photon-photon annihilation just a  small fraction is attenuated.

The proton luminosities were normalized by assuming the hadronic emission exhausts the observed GeV-TeV gamma-ray flux. Requiring the BH mass for Mrk 421 of $2\times 10^8 M_\odot$ \citep{2003ApJ...583..134B},  the proton luminosities are less than the Eddington luminosity $L_{Edd}= 2.5\times 10^{46}\,{\rm erg\, s^{-1}}$.   The emitting radii found are in the range of $\sim 10^{14} - 10^{16}$ cm for the both scenarios, which are larger than the gravitational radius $r_g=5.9\times 10^{13}$ cm.    The minimum Lorentz factors for the proton population were calculated following the charge neutrality condition; a comparable amount of electrons and protons.  From the values of the magnetic field, electron, and proton luminosities, we can observe that a principle of energy equipartition $\lambda_{kl}=\frac{L_k}{L_l}$ could be present in this flaring event. Tables \ref{sed_parameters_hadronic_one} and \ref{tab_parameters_twozone_hadronic} indicate that comparable results may be obtained for both lepto-hadronic models. 

The values exhibited  in Table \ref{sed_parameters_leptonic} are in the range of those reported by  \cite{2011ApJ...736..131A} and \cite{2015A&A...578A..22A}.   For instance, the values of the Doppler factor, emitting radius and the magnetic field are $\approx 24$, $\approx 3.2\times 10^{16}\,{\rm cm^{-3}}$, and $45 {\rm mG}$, respectively.  The minimum variability timescale becomes  $\tau_{\rm \nu, min} \approx 0.5\,{\rm days}$.  The large value of $\gamma_{\rm e, min}$ required in this model to describe the SED implies that the Fermi mechanism efficiently accelerates electrons.  Given the value of the magnetic field, the maximum Lorentz factor of the electron distribution becomes $\approx 2\times 10^8$.  By comparing the values of $\gamma_{\rm e}$ with $\gamma_{\rm e, c1}$ and $\gamma_{\rm e, c2}$, it can be observed that the second break in the electron population, as defined by the double-break PL, exhibits affinity to the second-break Lorentz factor ($\gamma_{\rm e, c2}$). This similarity is likely attributed to the influence of synchrotron cooling. According to \cite{2011ApJ...736..131A}, the Lorentz break $\gamma_{\rm e, c1}$ corresponds to the acceleration mechanism, with a decrease in the efficiency of electron acceleration observed for energetic electrons above this break.



\section{Neutrino Analysis and Description}\label{neu_event}


\subsection{Data Analysis}

In order to establish potential coincidences between the IceCube neutrinos and their corresponding electromagnetic counterparts, a temporal and coincidence correlation was performed. This correlation aimed to identify matches for each neutrino mentioned in the HESE catalog and Mrk 421 around the 2012 and 2013 flaring activities.   We consider the angular separations and the uncertainties associated with the neutrino's positions and arrival times. We found one potential correlation (event IC31)  between IceCube neutrinos and the source Mrk 421. The neutrino event IC31 listed in the HESE catalog was detected in temporal coincidence with F5.   This neutrino shower-like event detected on 2012 September 06 had an energy of $42.5^{+5.4}_{-5.7}\,{\rm TeV}$ centered around the coordinates  ${\rm RA= 146.1^\circ}$ and ${\rm Dec= 78.3}$ (J2000). The spatial correlation between this astrophysical neutrino event and the observed electromagnetic flare is obtained considering that this is a shower-like event with $26^{\circ}$ of median angular resolution and the position of the BL-Lac Mrk 421 is at 1.5 $R_{50}$\footnote{the positional uncertainties called $R_{50}$ refers to an area of the sky delimited by the angular resolution obtained for the reconstructed event. The introduction of this parameter underlines that only half of all reconstructed events can originate inside the measured $R_{50}$ regions while the other half can come from larger offset angles.}.  The features of the event IC31 are reported in Table \ref{table:neutrino}.

\subsection{Number of Neutrino events}
Accelerated protons interact with a low-energy bump by p$\gamma$ interactions.   Photo-hadronic interactions in the dissipation zone also generate neutrinos through the charged pion decay products ($\pi^{\pm}\rightarrow \mu^\pm+  \nu_{\mu}/\bar{\nu}_{\mu} \rightarrow  e^{\pm}+\nu_{\mu}/\bar{\nu}_{\mu}+\bar{\nu}_{\mu}/\nu_{\mu}+\nu_{e}/\bar{\nu}_{e}$).   Following \citet{2014PhRvD..90b3007M}, the neutrino spectrum is estimated through the proton spectrum given by 
\be\label{pg}
\epsilon_\nu L_\nu\simeq f_{\pi^0} \varepsilon_p L_p\,,
\ee
where $f_{\pi^0}$ is given in Eq. \ref{eficiency} considering that each neutrino and photon bring 5\% and 10\% of the initial proton energy, respectively.   It is worth noting that photon and neutrino spectra  from pion decay products are associated by {\small $\int \left(dN/d\epsilon\right)_\nu\,\epsilon_\nu\,d\epsilon_\nu=\frac14\int \left(dN/d\epsilon\right)^{\pi^0}_{\gamma}\,\epsilon^{\pi^0}_{\gamma}\,d\epsilon^{\pi^0}_{\gamma}$}  \citep{2008PhR...458..173B}. Therefore, the characteristic neutrino energy is estimated from Eq. \ref{pgamma} as $\epsilon_{\rm \nu,br}\approx \epsilon^{\pi^0}_{\gamma,c}/2$. 

The number of neutrino events ($N_{\rm ev}$) expected in the IceCube telescope with a neutrino flux $\left(dN/d\epsilon\right)_\nu$ and a time of observation ($T$) can be derived through the following relation
\bary\label{evtrate}
N_{\rm ev}&\simeq &T\,\int_{\epsilon_{\rm \nu, th}} A_{\rm eff}(\epsilon_{\rm \nu})\, \left(\frac{dN}{d\epsilon}\right)_\nu \,d\epsilon_{\rm \nu},
\eary
where $\epsilon_{\rm \nu, th}$ is the threshold energy  and  {\small $A_{eff}$} is the effective area available in the IceCube data release webpage \footnote{https://icecube.wisc.edu/science/data-releases/}.


%
\subsection{Analysis of astrophysical neutrino signal  within the time window of the flare}
%
%
A promising analysis to associate the astrophysical neutrino events with known sources is the time correlation with the significant observed gamma-ray flares \citep{2016arXiv160202012K}.   This is the case of the IceCube event IC31 if we consider the major gamma-ray flare measured by Fermi-LAT and ARGO-JBJ experiments started in July 2012 and finished in September of the same year.  Considering the one- and two-zone lepto-hadronic scenarios used to describe the gamma-ray flare {\bf and using Eq. \ref{pg}}, we extrapolate the expected neutrino flux from Mrk 421 during this high activity period. {\bf It is worth noting that the values of the Doppler factors found in the range of $4 - 17$ and the seed photons at ${\sim \rm keV}$ energies lead to neutrino events of dozens of TeV. The corresponding neutrino event rate expected in the IceCube observatory in different time windows is reported in Table \ref{table:fluxFermi}. This estimate is obtained following the procedure explained in the previous subsection.   The flaring activity of Mrk 421 detected in GeV gamma-rays and radio wavelengths lasted $\sim 100\,{\rm day}$, so the most relevant time window corresponds to this period. In this case, the neutrino event expected for one- and two-zone lepto-hadronic model is 0.11 and 0.13, respectively.  It is seen that the lepto-hadronic models exhibit degeneracy with the parameters used, meaning that by using a totally different set of parameters, identical outcomes may be achieved.   Hence, our outcome is not exclusive, but rather represents just one potential solution to model these flare activities.   The broadband SEDs are directly affected by the MeV gamma-ray flux produced by synchrotron radiation of secondary pairs (e.g., see Figure \ref{fig_SED_flare_2012_two}), so Mrk 421 might be studied by MeV gamma-ray orbiting observatories like All-sky Medium Energy Gamma-ray Observatory  (AMEGO) \citep{2019BAAS...51g.245M} and enhanced ASTROGAM (e-ASTROGAM) \citep{2018JHEAp..19....1D}, which  will survey the sky in the energy ranges of 0.3–100 MeV and 0.3–3 GeV, respectively. Therefore, detecting gamma-ray flux could constrain the parameters and estimate the neutrino events.  For the IceCube event IC31, the corresponding very-high energy (VHE) photons produced through the p$\gamma$ interactions  would have energies around $\rm \sim 100\, {\rm TeV}$.   When traveling between Mrk 421 and Earth, the VHE flux around these energies would be severely weakened by the EBL effect. For instance,  considering the EBL model proposed by \cite{2017A&A...603A..34F}, the absorption multiplicative factor to the VHE flux becomes $e^{\rm -\tau_E(z, E_\gamma)}$ with the photon-photon optical depth ${\rm \tau_E(0.031, 100\, TeV)} \sim 50$. Given the features of broadband SED at keV energies, Mrk 421 is a blazar usually classifies as a high-peaked synchrotron source \cite{2010ApJ...716...30A}. In this case, neutrino events of dozens of TeV are expected. If the relativistic protons accelerated in the emitting region interact with the seed photons at eV energies (i.e., the lower energy peak in the broadband SED), neutrino events of a few PeV might be expected, which could be the case of the low- and intermediate-peaked synchrotron sources.}


Aartsen et al. 2013 used data collected during the initial period of operations (from June 2010 to May 2011) of IceCube's DeepCore low-energy extension to perform the first measurement of the atmospheric electron neutrino flux in the energy range spanning from 80 GeV to 6 TeV \cite{2013PhRvL.111h1801A}.   In 281 days of data, an observational sample of 1029 events is obtained, of which ${\rm 496\pm 66(stat)\pm 88(syst)}$ are assessed to represent cascade events. These estimates include both electron neutrino and neutral current event observations.  The observed flux of electron neutrinos in the atmosphere aligns with theoretical predictions for atmospheric neutrinos within this specific energy range \cite{2007PhRvD..75d3006H, 2004PhRvD..70b3006B}. The best-fit fluxes for conventional electron and muon neutrinos becomes $\epsilon^2_{\nu_e} (dN/d\epsilon)_{\nu_e}=(1.44\pm0.52)\times 10^{-2}\left(E/{\rm 1\,GeV}\right)^{-3.21\pm0.68}\,{\rm GeV\, cm^{-2}\, s^{-1}\,sr^{-1}}$  and $\epsilon^2_{\nu_\mu} (dN/d\epsilon)_{\nu_\mu}=(2.09\pm0.77)\times 10^{-2}\left(E/{\rm 1\,GeV}\right)^{-3.23\pm0.05}\,{\rm GeV\, cm^{-2}\, s^{-1}\,sr^{-1}}$, respectively. Using the Eq. \ref{evtrate} and extrapolating the neutrino fluxes up to dozens of TeV, we estimate the number of atmospheric neutrinos, as listed in Table \ref{table:fluxFermi}.  This Table shows that that the number of atmospheric neutrinos is far below the neutrino events estimated for Mrk 421.



\section{Ultra-high-energy cosmic rays}\label{uhecr}


\subsection{Data Analysis}

We perform temporal and spatial correlations between the TA events and Mrk 421 around the 2012 and 2013 flaring activities.  We use the list of 72 events with $E>57\,{\rm EeV}$ that have been collected by surface detector of the TA experiment from 2008 May 11 to 2013 May 04 \cite{2014ApJ...790L..21A}.  Ryu et al. (2010) estimated that the deflection angle between the arrival direction of UHE protons and the sky position is quite large with a mean value of $<\theta_d>\simeq 15^\circ$ \cite{2010ApJ...710.1422R}. In addition, Das et al. (2008) found that for observers situated within groups of galaxies like ours, about 70\% (35 \%) of UHECRs with energies above 60 EeV emerge inside $\sim 15^\circ$ ($5^\circ$), of the astrophysical object position \cite{2008ApJ...682...29D}.  In order to find potential coincidences of UHECRs with Mrk 421 we consider an angular separation of $15^\circ$ of each event.  We found three events, two associated with the flare in 2012 (F5) and one with the flare in 2013 (F6). The features of these events are listed in Table \ref{table:uhecr}.





\subsubsection{Deflection Angle and Time delays}

VHE $\gamma$-ray blazars with low redshifts have been proposed as suitable sources for researching the UHECR fluxes \citep{2012ApJ...749...63M, 2009NJPh...11f5016D, 2012ApJ...745..196R,2010ApJ...719..459J, 2021PhRvL.126s1101R}.  Therefore,   UHECR activities on timescales as long as days, weeks, or months could be associated with the $\gamma$-ray activities in blazars.  However,   UHECRs traveling from the source to Earth randomly deviate between the original and the observed arrival direction by the strength of intergalactic and Galactic magnetic fields \citep{2008ApJ...682...29D, 2010ApJ...710.1422R, 2016MNRAS.456.1723L}. 
\paragraph{Case (i) Galactic magnetic field.} It is possible to model the Galactic magnetic field as a magnetic disk of a certain height ($h_{\rm md}$), resulting in a deflection angle of $\theta_{\rm dfl}\approx \frac{h_{\rm md} \csc b}{r_L}$, where  $r_L=\frac{E_{\rm Z}}{ZeB}$ is the Larmor radius and {\rm b} is the Galactic latitude of Mrk 421. In this case, the Larmor radius and the deflection angle for nuclei with energy $E_{\rm Z}$ and charge Z are

\be
r_{\rm L,G}\simeq 107\,{\rm kpc}\, Z^{-1}\left(\frac{E_{\rm Z}}{100\,{\rm EeV}} \right) \left(\frac{B_{\rm G}}{10^{-6}\,G} \right)^{-1}\,,
\ee
and
\be
\theta_{\rm d, G} \lesssim 0.6^\circ \, \frac{Z}{\sin b}\left(\frac{h_{\rm md}}{1\,{\rm kpc}}\right) \left(\frac{E_{\rm Z}}{100\,{\rm EeV}} \right)^{-1} \left(\frac{B_{\rm G}}{10^{-6}\,G} \right)\,,
\ee

respectively, restricted by the size of the magnetic disk, which is finite. For instance,  magnetic field averages in the thick gaseous disk of our Galaxy $\approx 0.2\,{\rm kpc}$ lie in the range of $\approx (3 -5)\times 10^{-6}\,{\rm G}$, but might be much less $\ll 1\times 10^{-6}\,{\rm G}$ in the kpc-scale halo \citep{2002ApJ...572..185A}.   

The time delay between the photons and UHECR events  coming from the flaring activity can be calculated by
{\small
\bary\label{time_delay_Gal}
\Delta t &\approx& \frac{h^3_{\rm md}}{24\,r^2_L\,}\cr
&\approx& 8.3\, {\rm day}\,Z^2 \left(\frac{h_{\rm md}}{1\,{\rm kpc}} \right)^3\left(\frac{E_{\rm Z}}{100\,{\rm EeV}} \right)^{-2}\left(\frac{B_{\rm G}}{10^{-6}\,G} \right)^2.
\eary
}

\paragraph{Case ii) Intergalactic magnetic field.} The deflection angle due to the intergalactic magnetic field could be estimated as \citep{1995ApJ...452L...1W, 1995PhRvL..75..386W, 2009NJPh...11f5016D,
2021PhRvD.104h3017A, Farrar_2014CRPhy..15..339F, Kievani_2015APh....61...47K, 2017PhRvD..96b3010A, 2005JCAP...01..009D,
 2022arXiv221013052X, 1995Natur.374..430P, 2017ApJ...847...39V, 1999ApJ...514L..79B}
{\small
\bary
\theta_{\rm d, IGM}&\simeq& \frac{d_z}{2r_L\sqrt{N_{\rm inv}}}\cr
&\simeq&0.01^\circ\,Z \left(\frac{d_z}{100\,{\rm Mpc}} \right)\left(\frac{E_{\rm Z}}{100\,{\rm EeV}} \right)^{-1}\left(\frac{B_{\rm IGM}}{10^{-12}\,G} \right)\cr
&& \hspace{3.8cm}\times\left(\frac{N_{\rm inv}}{10^2} \right)^{-1/2}.
\eary
}
where $N_{\rm inv}\simeq {\rm max(\frac{d_z}{\lambda}, 1)}$ is the number of reversals of the magnetic field, $\lambda$ is the magnetic-field correlation length and $d_z$ corresponds to the distance between Earth and the source.  The time delay between the photons and UHECR events  coming from the flaring activity can be calculated by
{\small
\bary\label{time_delay_int}
\Delta t &\approx& \frac{d^3_z}{24\,r^2_L\,\sqrt{N^3_{\rm inv}}}\cr
&\approx& 4.6\, {\rm day}\,\,Z^2 \left(\frac{d_z}{100\,{\rm Mpc}} \right)^2\,\left(\frac{E_{\rm Z}}{100\,{\rm EeV}} \right)^{-2}\left(\frac{B_{\rm IGM}}{10^{-12}\,G} \right)^2\cr
&& \hspace{4.6cm}\times
\left(\frac{N_{\rm inv}}{10^2} \right)^{-3/2}.\,\,\,
\eary
}


It is worth noting that the maximum time delay (Eqs. \ref{time_delay_Gal} and \ref{time_delay_int}) due to the Galactic and intergalactic magnetic field is at a timescale of months.

\subsubsection{Number of UHECRs and Maximum Energy}
Requiring that a supermassive black hole (BH) around Mrk 421 has the power to accelerate particles up to UHEs through Fermi processes,  protons accelerated in the blazar zones are confined by the Hillas condition \citep{1984ARA&A..22..425H}.    During the flaring intervals for which the apparent isotropic luminosity can reach $\approx 10^{45}$ erg s$^{-1}$ and from the equipartition magnetic field $\epsilon_B$,  the maximum particle energy of accelerated UHECRs depends on the region size and on the assumed acceleration efficiency which can be written as \citep{2009NJPh...11f5016D}

\be
\varepsilon_{\rm Z, max}\approx 10^{20}\,\frac{Z\eta_{\rm acc}q_e\,\epsilon^{1/2}_B}{\, \Gamma}\,\left(\frac{L_p}{10^{45}\,{\rm erg/s} }\right)^{1/2}\, {\rm eV}\,,
\ee
where $Z$ is the atomic number.  To estimate the number of UHECRs,  we take into account the TA  exposure,  which  for a point source is given by $E_{\rm TA}  \omega(\delta_s)$, where  $\omega(\delta_s)$ is an exposure correction factor for the declination of Mrk 421 \citep{2001APh....14..271S, 2012NIMPA.689...87A, 2016ApJ...830...81F}.  However, when considering p-gamma losses as estimated in Section \ref{sec:2} for one- and two-zone lepto-hadronic model the maximum particle energy is recalculated.   In the two-zone model, protons are accelerated in the inner blob, and to estimate the maximum proton energy reached  inside the inner blob,  we compare the acceleration timescale, $t_{\rm acc}' \simeq \varepsilon_p'/(e B_i' \eta_{\rm acc})$,  with losses timescales, ${\rm min}\{t_{\rm ad}, t_{\pi}\}$. In this case, the adiabatic loss and photopion timescales are of the order of dynamical timescale, $t_{\rm ad}' \simeq t_{\rm dyn}' \sim 3.3 \times 10^3 \, {\rm s} \, (R_i'/10^{14}\rm \, cm) $, and $t_{\pi}' \sim \sigma_{\gamma\gamma}/ (\sigma_{p\pi}\tau_{\gamma\gamma}) t_{\rm dyn}' \sim 2 \, t_{\rm dyn}'$, respectively, where $\sigma_{\gamma\gamma}/\sigma_{p\pi} \sim 266$ and $\tau_{\gamma\gamma} \sim 133$. Therefore, in the comoving frame the maximum energy that protons can reach is estimated by $ \varepsilon_{p, \rm max}' \sim 2.7 \, {\rm EeV \, } \eta_{\rm acc} \, \left( B_i'/100 \, \rm G\right) \left( R_i' /10^{14} \, \rm cm\right)$. We use this value to model the emission during flare states. Similarly, we estimate the maximum energy for one-zone leptohadronic model.   Taking the parameter values listed in Table \ref{sed_parameters_hadronic_one}, the maximum energy that protons can reach during 2012 flare is $ \varepsilon_{p, \rm max} \sim 72.6 \, {\rm EeV \, }  \eta_{\rm acc}  \, \left( B'/19 \, \rm G\right) \left( R_b' /7.5\times 10^{14} \, \rm cm\right) \left(\Gamma /17\right)$, and during 2013 flare is  {\small $ \varepsilon_{p, \rm max} \sim 72.8 \, {\rm EeV \, } \eta_{\rm acc} \, \left( B'/20 \, \rm G\right) \left( R_b' /8.1\times 10^{14} \, \rm cm\right) \left( \Gamma /15\right) \,$}  Taking the parameter values listed in Table \ref{sed_parameters_leptonic} (one - zone SSC scenario), the maximum energy that protons can reach during 2013 flare is {\small$ \varepsilon_{p, \rm max} \sim 10.3 \, {\rm EeV \, } \eta_{\rm acc} \, \left( B'/22.6\times 10^{-3} \, \rm G\right) \left( R_b' /6.4\times 10^{16} \, \rm cm\right) \left( \Gamma /24\right) \,$} The values of the one-zone scenario are in agreement with those reported in \cite{2013SAAS...40..225D}. 
 
 We estimate the expected number of UHECRs in TA experiment as 
\bary
N_{\tiny UHECRs}&=& {\rm F_r\, E_{TA}\,\omega(\delta_s) \,\frac{F_{\rm EeV}}{\varepsilon_{\rm EeV}}}\,,
\label{num}
\eary
where $F_r\simeq 0.25$ is the fraction of propagating cosmic rays that survives over a distance $d_z$ \citep{2011ARA&A..49..119K}  and  $F_{\rm EeV}$  is the flux of UHECRs, which is estimated extrapolating the proton spectrum up to ultra-high energies $F_{\rm EeV}=\frac{L_p}{4\pi d_z^2}\left(\frac{\varepsilon_{\rm EeV}}{\varepsilon_{\rm GeV}}\right)^{-\alpha_p+2}$, $\varepsilon_{\rm EeV}$ corresponds to the energy of the UHECR event recorded by TA experiment and $\varepsilon_{\rm GeV}$ to the average energy for describing the LAT observations $\sim 100\,{\rm GeV}$. The values of $L_p$ and $\alpha_p$ are obtained from the parameters reported in Tables \ref{sed_parameters_hadronic_one} and \ref{sed_parameters_leptonic}.   Taking into account the TA exposure  (3690 km$^2$ sr yr) \cite{2013ApJ...768L...1A} and the parameters reported in Tables \ref{sed_parameters_hadronic_one}, \ref{tab_parameters_twozone_hadronic} and \ref{sed_parameters_leptonic}, from Eq. (\ref{num}) we found that the number of UHECRs lies in the range of 0.66 - 1.89. {\bf Considering the acceleration efficiency factor of $\eta_{\rm acc}\sim 0.2$, the time delay between photons and UHECRs of a timescale of months and the deflection angle due to Galactic and intergalactic magnetic fields $\lesssim 15^{\circ}$}, we note that although an analysis of chemical composition must be performed  nuclei with $Z\sim 4$ are more favorable to be expected from this source. Therefore, comparing the results obtained from our model with the those UHECRs reported by the TA experiment, it can be seen that these events could be associated with Mrk 421 during the flaring activities.
\section{Conclusions}
%
%
We have used one- and two-zone lepto-hadronic models to characterize and describe the broadband SEDs of Mrk 421 in light of the fact that: i) the flaring activity between July 2012 and September 2012 produced the most significant GeV gamma-ray flux ever observed  with low (absence) activity in soft (hard) X-rays, and ii) the flaring activity in April 2013 exhibited the strongest flux ever recorded in TeV gamma-ray and optical bands.  We have shown that accelerated protons interacting with keV photons and suitable values of parameters could give an account of high flux at the GeV-TeV energy bands with different levels of flux at the X-ray band. In addition, we have modelled the flaring activity in 2013 April with one-zone SSC scenario. The entire sets of parameters obtained using the one-zone lepto-hadronic and SSC model are consistent with those reported in \cite{2011ApJ...736..131A, 2015A&A...578A..22A} and those values found using the two-zone lepto-hadronic model with previous descriptions of blazars \citep{AguilarRuiz_2022MNRAS.512.1557A, AguilarRuiz_2023JHEAp..38....1A}. 

We performed a temporal and spatial correlations among the neutrino events reported by IceCube in the HESE catalog, the UHECR events collected by surface detector of the TA experiment, and Mrk 421 around the 2012 and 2013 flaring activities. For the neutrino events we consider the angular separations and the uncertainties associated with the neutrino's positions, and for UHECR events we consider an uncertainty of $15^\circ$ around the UHECR's positions.  We found that the neutrino shower-like event IC31 detected on 2012 September 06 with an energy of $42.5^{+5.4}_{-5.7}\,{\rm TeV}$ and two UHECRs detected on 2012 September 07 and  2012 September 27 with energies of $57.8$ and $57.4$ EeV, respectively, were in temporal coincidence with F5, and one UHECR detected on 2013 April 22 with energy of $62.5$ EeV  with F6.  We have calculated the number of neutrino event rates and the atmospheric neutrino background predicted in the IceCube neutrino observatory during the flare time by correlating the measured GeV-TeV gamma-ray flux and the expected HE neutrino flux through p$\gamma$ interactions.    We have estimated the maximum energy that CRs can be accelerated in the relativistic jet and the number of UHECRs that could be expected from Mrk 421. The analysis showed that the estimated number of UHECRs lies in the range of 0.66 - 1.89, so that UHECRs reported by the TA experiment could be associated with Mrk 421 during the flaring activities.

\section*{Acknowledgements}
We express our gratitude to the anonymous referee for their meticulous examination of the paper and insightful recommendations, which greatly enhanced the excellence and lucidity of our manuscript.  We also thank to Markus Adler, Antonio Marinelli, Francis Halzen and  Ignacio Taboada for useful discussions, Atreyee Sinha and Chen Songzhan  for sharing with us the data.  NF acknowledges financial support  from UNAM-DGAPA-PAPIIT  through  grant IN106521. 

\clearpage
\bibliographystyle{elsarticle-harv} 
\bibliography{main}
\addcontentsline{toc}{chapter}{Bibliography}
\clearpage

\begin{table*}
\begin{center}\renewcommand{\arraystretch}{1.3}\addtolength{\tabcolsep}{1pt}
\caption{Integrated flux in the 0.3-100 GeV energy range around the flaring activity of 2012, considering a PL spectral index $\alpha_{\gamma, LAT}=1.75$ \citep{2016ApJS..222....6B}}\label{table:gammaflux}
\begin{tabular}{ c c }
  \hline \hline
 \normalsize{$T$} & \hspace{0.5cm}\normalsize{$F_{\gamma}$}  \\
 \normalsize{(days)} & \hspace{0.5cm}\normalsize{(GeV$^{-1}$ cm$^{-2}$ s$^{-1}$)} \\
 \hline
 \hline
\normalsize{$20 $} & \hspace{0.5cm}\normalsize{$9.9\times 10^{-8}$}    \\
\normalsize{$ 60$} & \hspace{0.5cm}\normalsize{$8.5\times10^{-8}$}     \\
\normalsize{$100$} & \hspace{0.5cm}\normalsize{$6.5\times10^{-8}$} \\
 \hline
 \end{tabular}
\end{center}
\end{table*}

%
\begin{table*}
\begin{center}\renewcommand{\arraystretch}{1.3}\addtolength{\tabcolsep}{1pt}
\caption{Parameters obtained using the one-zone lepto-hadronic model during the 2012 and 2013 flaring activities of Mrk 421.}\label{sed_parameters_hadronic_one}
\begin{tabular}{ l c c  c c c c c}
\hline
\hline
\normalsize{} &  & \normalsize{\bf 2012}&  \normalsize{}  &  & & \normalsize{\bf 2013}&  \normalsize{}\\
\normalsize{} &  \normalsize{July 09 to 21}&  &\normalsize{July 22 to} & \normalsize{} &  \normalsize{April 16}&  &\normalsize{April 17} \\
\normalsize{} &  \normalsize{}&  &  \normalsize{September 16} \\

\normalsize{} &  & \normalsize{}&  \normalsize{} \\
\hline
\hline
\normalsize{$\mathcal{D}$} & \normalsize{17} &  & \normalsize{16} & &\normalsize{16} &  & \normalsize{15}\\
\normalsize{$B'$\, (G)} & \normalsize{18}&   & \normalsize{19} & &\normalsize{21} &  & \normalsize{22}\\
\normalsize{$R'_b$ ($10^{14}$\, cm)} & \normalsize{7.3} &  & \normalsize{7.5}  & &\normalsize{8.1} &  & \normalsize{8.1}  \\
\normalsize{$N'_e$ ($10^{4}$\, cm$^{-3}$)} & \normalsize{$8.6$} &    & \normalsize{$8.3$} & &\normalsize{$9.2$} &  & \normalsize{$9.4$}\\
\normalsize{$K'_p\,(10^{5}\,\rm GeV^{-1} \, cm^{-3})$}& \normalsize{$7.6$} &  & \normalsize{$8.5$}   & &\normalsize{$8.4$} &  & \normalsize{$8.3$} 
\\
\normalsize{$\varepsilon_{\rm p,max}$ (${\rm EeV}$)}$^\dagger$ & \normalsize{72.6}&   & \normalsize{72.6} & &\normalsize{72.8} &  & \normalsize{72.8}
\\
\normalsize{$L_p$  ($10^{46}$\,erg\, $s^{-1}$)} & \normalsize{$1.8$ }&   & \normalsize{$1.7$} & &\normalsize{$2.0$} &  & \normalsize{$1.7$} \\
\normalsize{$\alpha_{\rm p}$} & \normalsize{2.22} &    & \normalsize{2.25} & &\normalsize{2.31} &  & \normalsize{2.29} \\
\normalsize{$L_e$  ($10^{44}$ erg\, $s^{-1}$)} & \normalsize{$7.2$}&  &  \normalsize{$5.2$} & &\normalsize{$8.3$} &  & \normalsize{$7.5$} \\
\normalsize{$\alpha_{\rm e,1}$} & \normalsize{2.22} &   &  \normalsize{2.25} & &\normalsize{2.28} &  & \normalsize{2.28}\\
\normalsize{$\alpha_{\rm e,2}$} & \normalsize{2.8} &   &  \normalsize{2.9} & &\normalsize{2.86} &  & \normalsize{2.91}\\

\normalsize{$\gamma_{\rm e,min}$ } & \normalsize{700}&   & \normalsize{$700$} & &\normalsize{$700$} &  & \normalsize{$700$}\\
\normalsize{$\gamma_{\rm e,c}$ ($10^4$)} & \normalsize{0.5}&  &  \normalsize{0.2} & &\normalsize{1.1} &  & \normalsize{0.9}\\
\normalsize{$\gamma_{\rm e,max}$ ($10^8$) } & \normalsize{1.0}&   & \normalsize{1.1} & &\normalsize{1.1} &  & \normalsize{1.0}\\
\normalsize{$L_B$  ($10^{44}$ erg\, $s^{-1}$)} & \normalsize{$1.9$}&   &  \normalsize{$2.0$} & &\normalsize{$2.8$} &  & \normalsize{$2.7$} \\

\normalsize{$U_B/(U_p+U_e)$ ($10^{-2}$)}   & \normalsize{$1.0$}&   & \normalsize{$1.1$} & &\normalsize{$1.3$} &  & \normalsize{$1.5$} \\

\hline
\hline
\end{tabular}
\end{center}
$^\dagger$The values reported of $\varepsilon_{\rm p,max}$ are given for $\eta_{\rm acc}=1$.\hspace{6cm}
\end{table*}
%


\begin{table*}
\begin{center}\renewcommand{\arraystretch}{1.3}\addtolength{\tabcolsep}{1pt}
\caption{Parameters obtained using the two-zone lepto-hadronic model during the 2012 and 2013 flaring activities of Mrk 421.}\label{tab_parameters_twozone_hadronic}
\begin{tabular}{l c c c c c c}
\hline
\hline
& 
\multicolumn{2}{c}{\textbf{2012}}
& \textbf{} &
\multicolumn{2}{c}{\textbf{2013}}
\\
  & \normalsize{July 09 to 21}
  & \hspace{1cm}\normalsize{ July 22 to }
  & \textbf{}
  & \normalsize{April 16}
  & \hspace{1cm}\normalsize{April 17}
  \\
  & \normalsize{}
  & \hspace{1cm}\normalsize{September 16}
  & \textbf{}
  & \normalsize{}
  & \hspace{1cm}\normalsize{}
  
\\
\hline
\normalsize{ Inner Blob }
\\
\hline
\normalsize{$\mathcal{D}_i$}
& 4	
& 4	
&& 3
& 3
\\
\normalsize{$B' \; \rm (G)$ }   
& \normalsize{100}  
& \normalsize{100}  
&& 100
& 80
\\

\normalsize{$R_i'$}($10^{14} \, \rm cm$) 
& \normalsize{1} 
& \normalsize{1} 
&
& \normalsize{1} 
& \normalsize{1} 
\\
\normalsize{$K'_p(\rm 10^{8} \, GeV^{-1} \, cm^{-3})$}
    & \normalsize{$1.9$} 
    & \normalsize{$0.1$}  
    &
    & \normalsize{$1.2$} 
    & \normalsize{$1.2$}  
\\
\normalsize{$\varepsilon_{p,\rm max} (\rm EeV) ^\dagger$}
    & \normalsize{11.0}     
    & \normalsize{11.0}     
    &
    & \normalsize{8.0}     
    & \normalsize{8.0}     
\\
\normalsize{$L_p\rm \; (10^{46} \, erg \, s^{-1})$}
    & \normalsize{$1.6$}  
    & \normalsize{$0.4$}   
    && \normalsize{$1.5$}  
    & \normalsize{$ 0.2 $}   
\\
\normalsize{$\alpha_p$}
& \normalsize{2.2}  
& \normalsize{2.0}  
&& \normalsize{2.0}  
& \normalsize{2.0}  
\\
\normalsize{$L_e \rm \; (10^{42} \, erg \, s^{-1})$}
    & \normalsize{$4.7$}  
    & \normalsize{$2.3$}  
    && \normalsize{$1.1$}  
    & \normalsize{$1.1$}  
\\
\normalsize{$\alpha_{\rm e,1}$}
& \normalsize{2.2}  
& \normalsize{2.0}  
&& \normalsize{2.0}  
& \normalsize{2.0}  
\\
\normalsize{$\alpha_{\rm e,2}$} 
& \normalsize{3.2} 
& \normalsize{3.0} 
&& \normalsize{3.0} 
& \normalsize{3.0} 
\\
\normalsize{$\gamma_{\rm e, min}$}  
    & \normalsize{$1$} 
    & \normalsize{$1$} 
    && \normalsize{$1$} 
    & \normalsize{$1$} 
\\
\normalsize{$\gamma_{\rm e, c}$}  
    & \normalsize{$10$} 
    & \normalsize{$5$} 
    && \normalsize{$5$} 
    & \normalsize{$6$} 
\\
\normalsize{$\gamma_{\rm e, max}$}  
    & \normalsize{$10^6$} 
    & \normalsize{$10^6$} 
    && \normalsize{$10^6$} 
    & \normalsize{$10^6$} 
\\
\normalsize{$L_B\rm \; (10^{42} \, erg \, s^{-1})$}
    & \normalsize{$5.9$} 
    & \normalsize{$5.9$} 
    && \normalsize{$1.9$} 
    & \normalsize{$1.2$} 
\\
\normalsize{$U_B/(U_p+U_e) \, (10^{-4})$}
& \normalsize{$3 $}   
& \normalsize{$15 $}   
&& \normalsize{$1.2 $}   
& \normalsize{$0.8 $}   
\\
\\
%
%
\hline
\normalsize{ Outer Blob }\\
\hline
\\
\normalsize{$\mathcal{D}_o$}
& \normalsize{10}  
& \normalsize{12}  
&
& \normalsize{14}  
& \normalsize{14}  
\\
\normalsize{$B'' \rm \; (G)$}
& \normalsize{0.56}  
& \normalsize{0.56}  
&
& \normalsize{0.26}  
& \normalsize{0.34}  
\\
\normalsize{$R_o''$ ($10^{15} \, \rm cm$)} 
& \normalsize{3.6} 
& \normalsize{3.6} 
& 
& \normalsize{32} 
& \normalsize{32} 
\\
\normalsize{$\gamma_{\rm e, min}$}
& \normalsize{$10$} 
& \normalsize{$10$} 
&
& \normalsize{$50$} 
& \normalsize{$50$} 
\\
\normalsize{$\gamma_{\rm e, c} \, (10^{4})$}  
    & \normalsize{$7$} 
    & \normalsize{$7$} 
    &
    & \normalsize{$3$} 
    & \normalsize{$3$} 
    \\
\normalsize{$\gamma_{\rm e, max} \, (10^{5})$} 
    & \normalsize{$2.6$} 
    & \normalsize{$3$} 
    &
    & \normalsize{$5$} 
    & \normalsize{$5$} 
\\
\normalsize{$\alpha_{\rm e,1}$} 
& \normalsize{2}  
& \normalsize{2}  
&
& \normalsize{1.8}  
& \normalsize{1.8}  
\\
\normalsize{$\alpha_{\rm e,2}$} 
& \normalsize{3.2} 
& \normalsize{3.2} 
&
& \normalsize{3.0} 
& \normalsize{3.0} 
\\
\normalsize{$L_B$} $\rm \; (10^{44} \, erg \, s^{-1})$
    & \normalsize{$0.2$}  
    & \normalsize{$0.2$}  
    &
    & \normalsize{$6.2$}  
    & \normalsize{$11$}  
\\
\normalsize{$L_e$} $\rm \; ( 10^{44} \, erg \, s^{-1})$
    & \normalsize{$1.9$} 
    & \normalsize{$1.9$} 
    &
    & \normalsize{$1.3$} 
    & \normalsize{$1.3$} 
\\
\normalsize{$U_B/U_e$}  
& \normalsize{0.1} 
& \normalsize{0.1} 
&
& \normalsize{4.9} 
& \normalsize{8.5} 
\\
\hline
\hline
\end{tabular}
\end{center}
$^\dagger$The values reported of $\varepsilon_{\rm p,max}$ are given for $\eta_{\rm acc}=1$.\hspace{6cm}
\end{table*}

\begin{table*}
\begin{center}\renewcommand{\arraystretch}{1.3}\addtolength{\tabcolsep}{1pt}
\caption{Parameters obtained using the one-zone SSC model {\bf during the 2013 flaring activity of Mrk 421.}}\label{sed_parameters_leptonic}
\begin{tabular}{ l c c c c }
\hline
\hline
\normalsize{} & \normalsize{2013 April 16}& \hspace{1cm} \normalsize{2013 April 17} \\
\normalsize{} & \scriptsize{}&  \scriptsize{} \\

\hline
\hline
\normalsize{$\mathcal{D}$} & \normalsize{24} & \normalsize{25} \\
\normalsize{$B'$ ( $10^{-3}$ G)} & \normalsize{$22.6$}&  \normalsize{$22.3$} \\
\normalsize{$R'_b$ ($10^{16}$ cm)} & \normalsize{$6.3$} & \normalsize{$6.4$}    \\
\normalsize{$N'_e$ (cm$^{-3}$)} & \normalsize{$0.72$} &   \normalsize{$0.73$}\\
\normalsize{$\varepsilon_{\rm p, max} \, \rm (EeV)$}$^\dagger$ & \normalsize{10.3} &  \normalsize{11.0}\\
\normalsize{$L_p$  (${10^{44}\,\rm erg\, s^{-1}}$ )} & \normalsize{$4.3$}&  \normalsize{$2.7$} \\
\normalsize{$\alpha_{\rm p}$} & \normalsize{2.10} &  \normalsize{2.11}\\
\normalsize{$L_e$  (${10^{43}\,\,\rm erg\, s^{-1}}$ )} & \normalsize{$8.9$}&  \normalsize{$10.1$} \\
\normalsize{$\alpha_{\rm e, 1}$} & \normalsize{$2.10$} &  \normalsize{$2.11$}\\
\normalsize{$\alpha_{\rm e,2}$} & \normalsize{2.77} &   \normalsize{$2.74$}\\
\normalsize{$\alpha_{\rm e,3}$} & \normalsize{$4.84$} &   \normalsize{$4.50$}\\
\normalsize{$\gamma_{\rm min}$} & \normalsize{$700$}&  \normalsize{$700$}\\
\normalsize{$\gamma_{\rm e, c1}\,\,(10^4)$} & \normalsize{$2.4 $}&  \normalsize{$2.1$}\\
\normalsize{$\gamma_{\rm e, c2}\,\,(10^7)$} & \normalsize{$1.7$}&  \normalsize{$1.8$}\\
\normalsize{$\gamma_{\rm e, max}\,\,(10^{11})$ } & \normalsize{$5.6$}&  \normalsize{$5.7$}\\
\normalsize{$L_B$  (${10^{42}\,\,\rm  erg\, s^{-1}}$ )} & \normalsize{$4.4$}&  \normalsize{$4.8$} \\
\normalsize{$U_B/(U_p+U_e)\,\,10^{-3}$}   & \normalsize{$1.6$}&  \normalsize{$1.5$} \\
\hline
\hline
\end{tabular}
\end{center}
$^\dagger$The values reported of $\varepsilon_{\rm p,max}$ are given for $\eta_{\rm acc}=1$.\hspace{6cm}
\end{table*}

\begin{table*}
\begin{center}\renewcommand{\arraystretch}{1.3}\addtolength{\tabcolsep}{1pt}
\caption{IceCube event \cite{2013Sci...342E...1I} detected in temporal coincidence with the 2012 flaring activity of Mrk 421}\label{table:neutrino}
\begin{tabular}{ c c c c c c c}
  \hline \hline
 \normalsize{ID}  & \hspace{0.3cm}\normalsize{Deposited Energy}        &  \hspace{0.3cm}\normalsize{Time}                     & \hspace{0.3cm}\normalsize{Declination } &    \hspace{0.3cm}\normalsize{RA}      & \hspace{0.3cm}\normalsize{Med. Ang. Resolution}& \hspace{0.3cm}\normalsize{Topology} \\
 \hspace{0.3cm}\normalsize{}       & \hspace{0.3cm}\normalsize{(TeV) }                           &  \hspace{0.3cm}\normalsize{(MJD)}                   & \hspace{0.3cm}\normalsize{(deg)}           &     \hspace{0.3cm}\normalsize{(deg)}  & \hspace{0.3cm}\normalsize{(deg)}                                 & \hspace{0.3cm}\normalsize{}\\
 \hline
 \hline
\normalsize{31}     & \hspace{0.3cm}\normalsize{$42.5_{-5.7}^{+5.4}$}    & \hspace{0.3cm}\normalsize{$56176$} & \hspace{0.3cm}\normalsize{$78.3$}         & \hspace{0.3cm}\normalsize{$146.1$} & \hspace{0.3cm}\normalsize{$26.0$}                      & \hspace{0.3cm}\normalsize{Shower} \\
 \hline
 \end{tabular}
\end{center}
\end{table*}

\begin{table*}
\begin{center}\renewcommand{\arraystretch}{1.3}\addtolength{\tabcolsep}{1pt}
\caption{Comparison between the expected number of signal events in IceCube, based on the model presented in {\bf Section \ref{neu_event}} and the atmospheric background. The expected number of atmospheric neutrinos was estimated using data collected in \cite{2013PhRvL.111h1801A}.}\label{table:fluxFermi}
\begin{tabular}{ c c c c c c}
  \hline \hline
 \normalsize{$T$} & \hspace{0.3cm}\normalsize{$E_{\rm th}$} &  
 \multicolumn{2}{c}{ \hspace{0.3cm}\normalsize{$N_{\rm ev}$} } & \hspace{0.3cm}\normalsize{$N^{\rm atm}_{\rm ev}$} \\
\normalsize{(days)} & \hspace{0.3cm} \normalsize{(TeV)} &  
\text{one-zone} & \text{two-zone} & \hspace{0.3cm}\normalsize{} &\hspace{0.3cm} \normalsize{} \\
 \hline
 \hline
\normalsize{$20 $} & \hspace{0.3cm}\normalsize{$ > 25$}    & \hspace{0.3cm}\normalsize{ $3.8 \times 10^{-2} $ } & $2.3\times 10^{-2}$ &\hspace{0.3cm}\normalsize{ $1.5 \times 10^{-13}$ }  \\
\normalsize{$ 60$} & \hspace{0.3cm}\normalsize{$ > 25$}    & \hspace{0.3cm}\normalsize{ $7.5 \times 10^{-2} $ } & $6.1\times 10^{-2}$  & \hspace{0.3cm}\normalsize{$4.5 \times 10^{-13}$}  \\
\normalsize{$100$} & \hspace{0.3cm}\normalsize{$ > 25$}    & \hspace{0.3cm}\normalsize{ $1.3\times 10^{-1}$} & $1.1\times 10^{-1}$ & \hspace{0.3cm}\normalsize{ $7.4 \times 10^{-13}$}  \\
 \hline
 \end{tabular}
\end{center}
\end{table*}

\begin{table*}
\begin{center}\renewcommand{\arraystretch}{1.3}\addtolength{\tabcolsep}{1pt}
\caption{List of TA events \citep{2041-8205-790-2-L21} recorded around the 2012 and 2013 flaring activities of Mrk 421.}\label{table:uhecr}
\begin{tabular}{ c c c c c}
  \hline \hline
 \normalsize{Date and Time}  & \hspace{0.3cm}\normalsize{Zenith angle} &  \hspace{0.3cm}\normalsize{$E_{\rm EeV}$} & \hspace{0.3cm}\normalsize{RA} & \hspace{0.3cm}\normalsize{Declination} \\
 \hspace{0.3cm}\normalsize{(UTC)}               & \hspace{0.3cm}\normalsize{(deg)} & \hspace{0.3cm}\normalsize{(EeV) } & \hspace{0.3cm}\normalsize{(deg)} & \hspace{0.3cm}\normalsize{(deg)} \\
 \hline
 \hline
\normalsize{2012 Sep 07 22:30:26} & \hspace{0.3cm}\normalsize{$38.79$}    & \hspace{0.3cm}\normalsize{$57.8$} & \hspace{0.3cm}\normalsize{$158.60$} & \hspace{0.3cm}\normalsize{$60.26$} \\
\normalsize{2012 Sep 27 13:51:05} & \hspace{0.3cm}\normalsize{$45.68$}    & \hspace{0.3cm}\normalsize{$57.4$} & \hspace{0.3cm}\normalsize{$159.75$} & \hspace{0.3cm}\normalsize{$35.56$} \\
\normalsize{2013 Apr 22 01:13:31} & \hspace{0.3cm}\normalsize{$36.20$}    & \hspace{0.3cm}\normalsize{$62.5$} & \hspace{0.3cm}\normalsize{$165.28$} & \hspace{0.3cm}\normalsize{$52.35$} \\
 \hline
 \label{tabnu}
 \end{tabular}
\end{center}
\end{table*}

%
%


%

%


%


%
%
%

%
%
%
%
%

\end{document}